\documentclass[%
 reprint,
 amsmath,amssymb,
 aps,
]{revtex4-2}

\usepackage{graphicx}
\usepackage{dcolumn}
\usepackage{bm}

\usepackage{hyperref}

\begin{document}

\setlength{\textfloatsep}{10pt}
\setlength{\abovecaptionskip}{10pt} 
\setlength{\belowcaptionskip}{10pt} 

\preprint{APS/123-QED}

\title{Traveling-Wave Parametric Amplifier with Passive Reverse Isolation}

\author{C. S. Kow}
\author{M.T. Bell}
 \email{Matthew.Bell@umb.edu}
\affiliation{Engineering Department, University of Massachusetts Boston, Boston, Massachusetts, USA}
\date{\today}

\begin{abstract}
Traveling-wave parametric amplifiers (TWPAs) have attracted much attention for their broadband amplification and near-quantum-limited noise performance. TWPAs are non-reciprocal by nature providing gain for forward-propagating signals and transmission line losses for backward traveling waves. This intrinsic non-reciprocity is insufficient to protect sensitive quantum devices from back-action due to noise from warmer amplification stages in practical systems, and thus necessitates the need for bulky cryogenic isolators. We present a multi-stage Traveling-Wave Parametric Amplifier (mTWPA) that addresses this limitation by achieving passive in-band reverse isolation alongside near-quantum-limited noise performance. The multi-stage architecture consists of two, mode conversion stages and a reflectionless high-pass filter which provides the passive isolation. Experimental measurements of a prototype mTWPA demonstrated 20 dB of forward gain across a 1.6 GHz bandwidth and greater than 35 dB of reverse isolation. Noise measurements indicate performance at 1.7 times the quantum limit. This demonstrates that the increased complexity of a multi-stage TWPA design does not lead to significant added noise. The designed distribution of gain across the stages is engineered to minimize internal amplifier noise at the input, and we propose further optimization strategies in redistribution of the gain between the stages. This level of isolation effectively mitigates noise from warmer amplification stages, matching the performance of conventional isolators. The mTWPA approach offers a scalable path forward for more efficient and compact quantum circuit readout systems.
\end{abstract}

\maketitle

\section{Introduction}
In low-temperature electronics, quantum-limited broadband amplifiers are essential for high-fidelity readout of weak signals. Superconducting travel-wave parametric amplifiers (TWPAs) have emerged as a leading technology due to their broadband operation and near-quantum-limited noise performance \cite{aumentado_rev,Roch_rev,delsing3wm,olegtwpa,siddiqitwpa1,siddiqitwpa2,twpascience,whitetwpa}. TWPAs have been successfully implemented in a range of experiments, including circuit quantum electrodynamics \cite{oliver_guide,qubit1,wallraff100qubit,wallraffmulti}, axion dark matter detection \cite{darkmatter1,darkmatter2,darkmatter3}, spin qubit measurements \cite{spin_qubit2}, broadband microwave squeezing \cite{Blaissqueeze1,oliversqueezing1,Rochsqueezing1,asym_squeezing,contvarentangle}, and microwave kinetic inductance detector (MKID) readout for astronomical imaging \cite{twpamkid1,twpamkid2}. TWPAs are typically the first stage in a cascaded amplification chain, followed by a high-electron-mobility transistor (HEMT) amplifier and additional room-temperature amplification \cite{wallraff100qubit}. Since total system noise is predominantly set by the first amplification stage \cite{pozar}, achieving quantum-limited noise performance in the TWPA is highly desired.

TWPAs facilitate signal amplification through a nonlinear transmission line. A weak signal at frequency \(\omega_s\) co-propagates with a strong pump at frequency \(\omega_p\), enabling parametric amplification via either three-wave mixing (3WM) \cite{delsing3wm,3wm_kin} or four-wave mixing (4WM) \cite{whitetwpa,olegtwpa,twpascience,bellkerr,bellpatent,belltwpa,lefthandedtwpa,Rochphotonic,Eomtwpa,contvarentangle,Rochrevkerr}, depending on whether the transmission line exhibits second- or third-order nonlinearity. Under phase matching conditions between traveling waves, these processes result in efficient signal gain and the generation of an idler at frequency \(\omega_i\) \cite{boyd2008nonlinear,agrawal2013nonlinear}.  

The nonlinear transmission medium can be realized using Josephson junction arrays \cite{twpascience,whitetwpa,Rochphotonic} or high-kinetic-inductance superconductors \cite{Eomtwpa,3wm_kin}, both of which allow for the implementation of 3WM or 4WM through appropriate biasing strategies. A critical challenge in TWPA design is maintaining phase matching between the interacting waves as they propagate along the transmission line  \cite{siddiqitwpa1,siddiqitwpa2}. To address this, dispersion engineering techniques have been employed, such as the introduction of band gaps or resonant features into the TWPA dispersion \cite{siddiqitwpa1,siddiqitwpa2,twpascience,Rochphotonic}. Such approaches modify the dispersion relation achieving optimal gain close to the pump in addition to adverse impedance matching conditions. Other approaches have proposed utilizing phase matching in left-handed Josephson junction transmission lines which utilizes the left-handed dispersion relation and eliminates the need for resonant structures \cite{lefthandedtwpa}. Relevant to this work are approaches that leverage transmission lines that incorporate asymmetric superconducting quantum interference devices (SQUIDs) which enable phase matching by modifying and inverting the third-order (Kerr) nonlinearity \cite{bellkerr,belltwpa,Rochrevkerr}. Such approaches dislocate the region of optimal gain in frequency from the pump and demonstrate improved impedance matching.

TWPAs are nonreciprocal devices that amplify waves that co-propagate with the pump. Backward propagating waves counter to the pump experience a significant phase mismatch, effectively suppressing first order parametric processes. The isolation exhibited by the TWPA for reverse-propagating signals are due to transmission line losses of the TWPA. Thermal noise from warmer HEMT amplifier stages at 4 K, can radiate through the TWPA, saturate TWPA parametric processes and cause back-action on the quantum device measured due to the limited isolation provided by the TWPA.

TWPAs are susceptible to backward-propagating waves resulting from non-negligible impedance mismatches at the ports. These mismatches cause reflections that can redirect a portion of the amplified signal and noise backward through the TWPA, leading to back-action on the quantum device measured \cite{aumentadotwpaiso,Rochtwpaiso}. While out-of-band noise can be effectively mitigated using band-pass filtering, in-band noise presents a more significant challenge, necessitating the use of isolators at the TWPA input. A conventional readout chain may incorporate up to three to four isolators, one at the input to the TWPA and two to three between the TWPA and HEMT amplifier \cite{wallraff100qubit}. These microwave components introduce insertion loss which limit the quantum efficiency of the readout and occupy substantial physical space at the mixing chamber stage of a dilution refrigerator. Since conventional isolators rely on large magnetic fields and ferrite materials to achieve non-reciprocity, \cite{pozar} their scalability is limited, making the development of alternative methods necessary for the continued development of prototype superconducting quantum computers.

There has been significant interest in developing non-reciprocal devices that do not rely on strong magnetic fields to replace isolators in measurement setups. However, few attempts have been made to realize non-reciprocal amplifiers, with existing implementations achieving narrow bandwidths of order 10 MHz for simultaneous gain and isolation \cite{amp_nonrecip1,amp_nonrecip2,amp_nonrecip3,amp_nonrecip4,amp_nonrecip5}. Recently, two experimental demonstrations have explored approaches to achieve broadband gain and reverse isolation in a TWPA \cite{aumentadotwpaiso,Rochtwpaiso}. Both approaches employ a forward pump that co-propagates with a forward signal, generating parametric gain through a phase-matched 3WM or 4WM process. To achieve reverse isolation, a second reverse-propagating pump is introduced, and through a 3WM process, up converts reverse propagating signals out of band. These methods have demonstrated sufficient gain and isolation over bandwidths of up to 500 MHz.

Challenges associated with these approaches are the necessity to finely tune both second- and third-order nonlinearities of the TWPA by requiring either a current bias or the application of an external magnetic field. The two counter-propagating pump frequencies and their respective power levels must be carefully adjusted to achieve simultaneous phase matching of two parametric processes. The experimental setup also necessitates the integration of directional couplers at both the input and output ports of the TWPA to facilitate the introduction of the two co-propagating pumps, along with additional filters to prevent pump leakage to the quantum device under measurement.

A challenge with isolation based on a parametric conversion process is that unwanted highly phase mismatched parametric processes are not completely suppressed. The need to phase match a 3WM or 4WM forward parametric gain processes at the same time puts a restriction on how well unwanted parametric processes can be suppressed. TWPAs designed with reverse isolation through parametric conversion introduce unwanted parametric conversion processes which have been shown to down-convert high-frequency spectral content to in-band, causing unintended back-action on the quantum device under test \cite{aps_twpaiso1,aps_twpaiso2}.

\begin{figure*}
\includegraphics[scale=0.7]{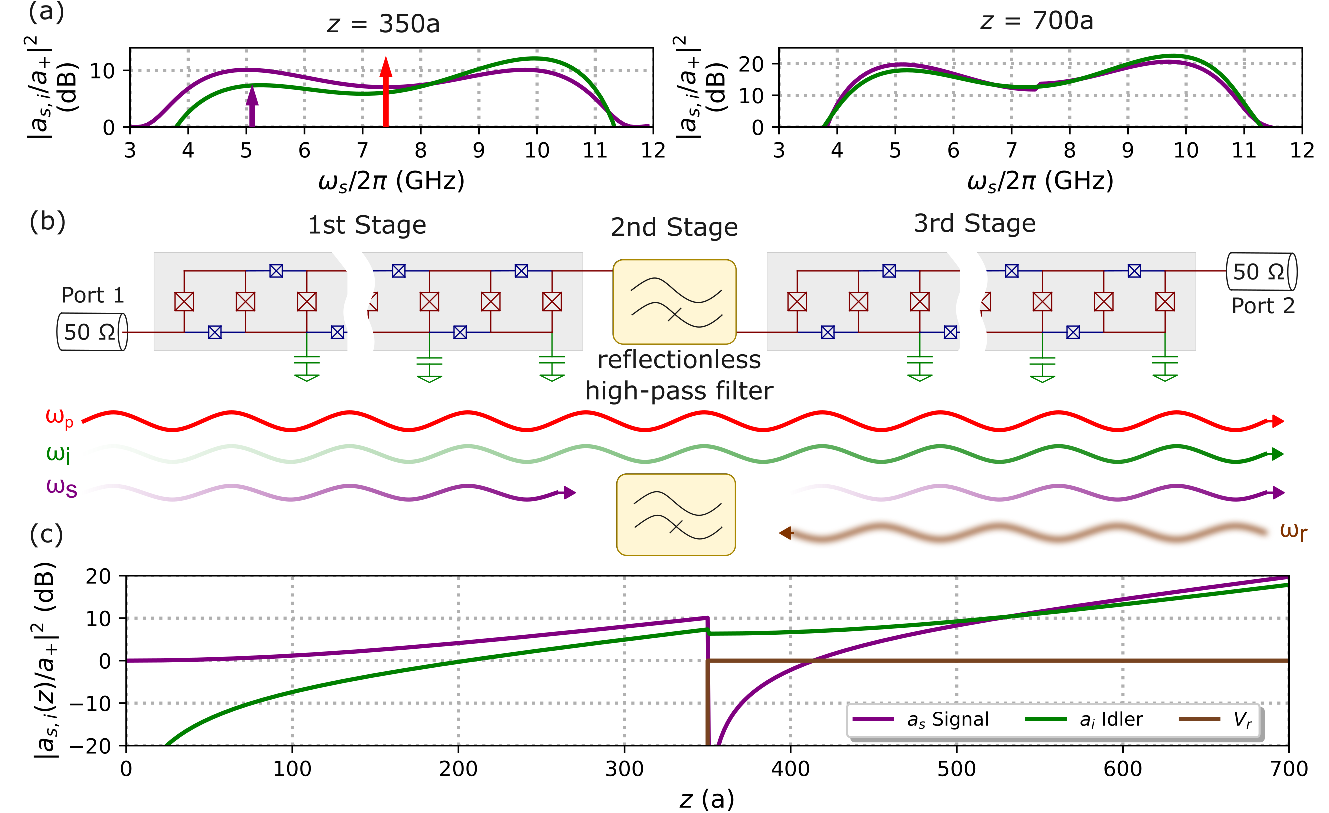}
\caption{\label{fig1}Signal propagation in mTWPA. (a) Transfer functions of the signal (purple line) and idler (green line) normalized to the input signal $|a_{s,i}/a_+|^2$ at the output of the first ($z = 350a$) and third stages ($z = 700a$). Position in frequency of input signal $\omega_s/2\pi = 5.2$ GHz and pump $\omega_p/2\pi = 7.4$ GHz indicated as purple and red arrows. (b) Three stage circuit diagram of the mTWPA. First and third stages of nonlinear transmission line composed of coupled asymmetric SQUIDs biased with an external magnetic flux. The second stage consists of a balanced high-pass reflectionless filter. Operation of the mTWPA consists of application of a pump (red line) and weak signal (purple line) at the input to the first stage. As a result of the 4WM process, an idler mode (green line) is generated in the first stage. In the second stage the weak signal is blocked by the filter and dissipated in a load instead of being reflected back into the first stage. The pump and idler tones propagate into the third stage where the signal mode is re-generated through the 4WM process. Backward propagating signals $\omega_r$ (brown line) are isolated by the second stage reflectionless filter. (c) Evolution of the signal (purple line) and idler (green line) along the length $z$ of the first and third stages of the mTWPA. Backward-propagating signal $\omega_r$ (brown line) isolated by the reflectionless filter.}
\end{figure*}

 The TWPA presented in this work is a multi-stage traveling wave parametric amplifier (mTWPA) that exhibits passive reverse isolation. The mTWPA, illustrated in Figure \ref{fig1} (b), consists of three distinct stages: (1) an input nonlinear transmission line parametric gain stage, (2) a reflectionless filter, and (3) an output nonlinear transmission line parametric gain stage. The first stage employs 4WM via a third-order nonlinearity and a strong pump at frequency $\omega_p$ to amplify a weak signal at frequency $\omega_s$ with an input amplitude $a_+$. The 4WM process is most effective when phase matching is satisfied, given by the condition $2k_p - k_s - k_i = 0$, where $k_n$ represents the wave number of each mode $n = \{s,i,p\}$ \cite{agrawal2013nonlinear,boyd2008nonlinear}. As a result of the 4WM process, an idler mode at frequency $\omega_i = 2\omega_p - \omega_s$ is generated.

This design utilizes an inverse Kerr phase matching technique \cite{belltwpa,bellkerr,bellpatent}, in which self-phase modulation (SPM) effects induced by the strong pump are countered by the chromatic dispersion of the transmission line. The transfer function of the first stage is depicted in Figure \ref{fig1} (a) for both the signal $|a_s/a_+(\omega_s)|^2$ (purple line) and the idler $|a_i/a_+(\omega_s)|^2$ (green line) modes. The positions in frequency of $\omega_s$ and $\omega_p$ are denoted by the purple and red arrows respectively. Figure \ref{fig1} (c) illustrates the evolution of $|a_{s,i}/a_+(z)|^2$ along the length $z$ across all three stages of the mTWPA in the phase-matching region where $\omega_s/2\pi = 5.2$ GHz. It is important that the phase matching region occurs away from the pump in frequency due to a finite roll-off of the stop-band of the reflectionless filter. Within the first parametric gain stage, $a_s$ and $a_i$ increase significantly with $z$, despite the idler mode initially being absent $a_i(0) = 0$ at the mTWPA input.

At the output of the first stage, the signal and idler modes enter the second stage, which functions as a reflectionless high-pass filter with a stop-band which includes $\omega_s$ and a pass-band encompassing both $\omega_p$ and $\omega_i$. Efficient 4WM processes in both the first and third parametric gain stages require that the TWPA transmission line be terminated in 50 $\Omega$ over a broad bandwidth that includes $\omega_s$ and $\omega_i$. This prevents the formation of standing waves that could otherwise disrupt the 4WM parametric process. The filter stage suppresses both the forward $\omega_s$ (purple line) and reverse $\omega_r$ (brown line) traveling waves within the stop-band by directing the undesired electromagnetic waves to be dissipated in one of two 50 $\Omega$ loads internal to the reflectionless filter. The reflectionless filter serves as the principal component providing passive reverse isolation in the mTWPA design.

The idler and pump propagate through the filter and enter the third stage of the mTWPA, where a second 4WM process amplifies $\omega_i$ and regenerates the signal tone at $\omega_s = 2\omega_p - \omega_i$. As depicted in Figure \ref{fig1}(c), $a_s$ at the output of the third stage surpasses $a_i$, and the transfer functions $|a_{s,i}/a_+(\omega_s)|^2$ at $z = 700a$ are shown in Figure \ref{fig1} (a). The experimentally realized mTWPA design demonstrates $20$ dB of forward gain and greater than $35$ dB of isolation for backward-propagating waves over a $1.6$ GHz bandwidth. This level of isolation is comparable to commercial double-junction cryogenic isolators \cite{isolator1,isolator2}, which are commonly used to attenuate noise from the HEMT amplifier stage in standard circuit QED cryogenic measurement setups.

\section{\lowercase{m}TWPA Design}
In the mTWPA design, the first and third parametric stages, Figure \ref{fig1}(b) employ an inverse Kerr phase-matching technique, which obviates the need for dispersion engineering and shifts the frequency of maximum gain away from \(\omega_p\). Given that the cutoff frequency $\omega_f$ of the reflectionless high-pass filter lies just below \(\omega_p\), and exhibits a finite roll-off response, the regions of maximum phase-matched gain in the parametric stages must be sufficiently displaced from $\omega_f$ to ensure a broad and usable bandwidth of forward signal gain and at the same time provide sufficient reverse isolation. The two-lobed gain profile, depicted in Figure \ref{fig1}(a), inherent to the inverse Kerr phase-matching technique \cite{belltwpa}, enables such displacement of phase matched gain from $\omega_p$.

The parametric stages of the mTWPA consist of a transmission line of coupled asymmetric SQUIDs \cite{bellkerr,bellsuperind,belltwpa}. The electrical circuit diagram is illustrated in Figure \ref{fig1}(b), where each unit cell comprises two coupled asymmetric SQUIDs of length \(a = 10\)~$\mu$m. One arm of the asymmetric SQUID contains a single small Josephson junction (denoted by the blue \(\times\)) with a critical current \(I_0\), while the other arm incorporates two larger Josephson junctions (denoted by the red \(\times\)) with a critical current of \(rI_0\), where \(r\) represents the ratio in the junction areas. The small and large junctions have shunting capacitance \(C_0\) and \(rC_0\), respectively. The SQUIDs are coupled through the large junctions, forming a meandered backbone (highlighted in red). To achieve a characteristic impedance of 50 \(\Omega\) in the transmission line, distributed parallel-plate capacitors to ground $C_{\rm gnd}$ are used. To minimize insertion loss, the dielectric material of the grounding capacitors is composed of low-loss SiN\(_x\). Each asymmetric SQUID is threaded by a magnetic flux \(\Phi\), enabling tunability of the mTWPA’s operating point. The total length of each transmission line for the first and third stages is denoted as $l$.

The propagation of electromagnetic waves in the first and third stages, with wavelength \(\lambda \gg a\), is governed by the wave equation:

\begin{eqnarray}\label{eq1}
\frac{1}{L}\frac{\partial^2\varphi}{\partial z^2}+C_0\left(\frac{r}{2}+2\right)\frac{\partial^4\varphi}{\partial t^2 \partial z^2}-C_{\rm gnd}\frac{\partial^2\varphi}{\partial t^2}-\nonumber\\
G\frac{\partial \varphi}{\partial t}-\gamma\frac{\partial}{\partial z}\left[\left(\frac{\partial \varphi}{\partial z}\right)^3\right]=0,
\end{eqnarray}

\noindent where $L=L_0/[r/2+2\cos(2\pi\Phi/\Phi_0)]$ and $\gamma=1/(3\varphi_0^2L_0)[r/16+\cos(2\pi\Phi/\Phi_0)]$ is the flux tunable linear inductance per unit-cell and Kerr constant respectively. $L_0=\Phi_0/\left(2\pi I_0\right)$ is the Josephson junction inductance,  $\Phi_0$ is the magnetic flux quantum, and $\varphi_0=\Phi_0/2\pi$. Losses in the parametric stages are introduced through a conductance $G=\omega C_{gnd} \tan\delta$ where $\tan\delta$ is the loss-tangent of the dielectric. The linear dispersion relation is

\begin{eqnarray}\label{eq2}
k(\omega, \Phi)=\frac{\omega\sqrt{L_0 C_{\rm gnd}}}{\sqrt{\left[\frac{r}{2}+2\cos\left(2\pi\frac{\Phi}{\Phi_0}\right)\right]-\omega^2L_0C_0\left(\frac{r}{2}+2\right)}}.
\end{eqnarray}

\noindent A solution to Eq. \ref{eq1} involves four traveling waves. In the degenerate case, this consists of two pump waves with equal angular frequencies \(\omega_p\), a signal \(\omega_s\), and a generated idler $\omega_i = 2\omega_p - \omega_s$. A set of coupled mode equations can be derived to describe the propagation and interaction of the signal and idler waves under the influence of a stiff pump:

\begin{eqnarray}\label{eq3}
\frac{\partial a_s}{\partial z}-i\kappa_s a_i^*e^{i\kappa z}=0,\nonumber\\
\frac{\partial a_i}{\partial z}-i\kappa_i a_s^*e^{i\kappa z}=0,
\end{eqnarray}

\noindent where $a_n$ is a complex amplitude of the mode $n = \{s,i,p\}$. The total phase mismatch is represented by $\kappa = \Delta k + \Delta k_{\rm NL}$ where $\Delta k_{\rm NL} = \alpha_s+\alpha_i-2\alpha_p$ is the mismatch due to SPM and cross-phase modulation (XPM) and $\Delta k = k_s + k_i-2 k_p$ is the chromatic dispersion of the transmission line. The SPM and XPM per unit length is:

\begin{eqnarray}\label{eq4}
\alpha_{s,i}=\frac{3\gamma k_{s,i}^2k_p^2|A_p|^2}{4C_{\rm gnd}\omega_{s,i}^2},\nonumber\\
\alpha_{p}=\frac{3\gamma k_p^5|A_p|^2}{8C_{\rm gnd}\omega_{p}^2},
\end{eqnarray}

\noindent where $A_p$ is the pump amplitude. The couplings between the signal and idler in Eq. \ref{eq3} are

\begin{eqnarray}\label{eq5}
\kappa_s=\frac{3\gamma k_p^2k_s k_i(2k_p-k_i)|A_p|^2}{8\omega_s^2C_{\rm gnd}}\nonumber\\
\kappa_i=\frac{3\gamma k_p^2k_s k_i(2k_p-k_s)|A_p|^2}{8\omega_i^2C_{\rm gnd}}.
\end{eqnarray}

\noindent The solution to the coupled mode equations \cite{siddiqitwpa2} described in Eq. \ref{eq3} are:

\begin{eqnarray}\label{eq6}
a_s(z) = \Bigg[ a_s (0) \left( \cosh gz - \frac{i\kappa}{2g} \sinh gz \right) +\nonumber\\ \frac{i\kappa_s}{g} a_i^* (0)  \sinh gz \Bigg]e^{i\kappa z/2}, \nonumber\\
a_i(z) = \Bigg[ a_i (0) \left( \cosh gz - \frac{i\kappa}{2g} \sinh gz \right) +\nonumber\\ \frac{i\kappa_i}{g} a_s^* (0)  \sinh gz \Bigg]e^{i\kappa z/2}
\end{eqnarray}

\noindent where $g$ is defined as:

\begin{eqnarray}\label{eq7}
g=\sqrt{\kappa_s\kappa_i-\left(\frac{\kappa}{2}\right)^2}.
\end{eqnarray}

\noindent From Eq. \ref{eq6}, the signal and idler power gains in the first and third stages are

\begin{eqnarray}\label{eq8}
H_{\rm s,i}=\cosh^2 gz + \frac{\kappa^2}{4g^2}\sinh^2 gz.
\end{eqnarray}

\noindent For the first stage of the mTWPA the power transfer function between a generated idler relative to an input signal for $a_i(0) = 0$ is

\begin{eqnarray}\label{eq9}
\left|\frac{a_i(z)}{a_s^*(0)}\right|^2=\frac{\kappa_i^2}{g^2}\sinh^2gz.
\end{eqnarray}

\noindent At the output of the first stage at $z=350a$, the transfer functions $|a_i/a_+|^2$ and $|a_s/a_+|^2$ as a function of $\omega_s$ is shown in Figure \ref{fig1}(a) as a solid green and purple line. As can be seen for signals with $\omega_s < \omega_p$ the idler does not grow to the same level as the signal. For $\omega_s > \omega_p$ the idler overtakes the signal.

For the third stage of the mTWPA the power transfer function between a generated signal relative to an input idler for $a_s(0)=0$,

\begin{eqnarray}\label{eq10}
\left|\frac{a_s(z)}{a_i^*(0)}\right|^2=\frac{\kappa_s^2}{g^2}\sinh^2gz.
\end{eqnarray}

\noindent At the output of the third stage at $z=700a$, the $|a_i/a_+|^2$ and $|a_s/a_+|^2$ as a function of $\omega_s$ is shown in Figure \ref{fig1}(a) as a solid green and purple line. These transfer functions represent the culmination of both stages in the mTWPA. The total power gain of the signal in the stop-band of the reflectionless filter between the first stage and the third stage absent loss is

\begin{eqnarray}\label{eq11}
G_s=\frac{\kappa_s^2\kappa_i^2}{g^4}\sinh^4gl.
\end{eqnarray}

Critical to achieving sufficient gain is the ability to tune the sign and magnitude of $\alpha_p$ through the flux tunable nonlinearity $\gamma$ and the pump amplitude $A_p$ to balance $\Delta k_{\rm NL}$ with $\Delta k$ and achieve phase matching with $\kappa \approx 0$ and $g^4 \approx \kappa_s^2\kappa_i^2$. For signals $\omega_s$ in both the pass- and stop-bands of the reflectionless filter under phase matching conditions, the signal gain can be approximated as $G_s \approx \exp (4\alpha_p l)/16$.

The second stage of the mTWPA is the high-pass reflectionless filter, which provides isolation in the stop-band of the filter. This filter is engineered to exhibit a stop-band at \(\omega_s\) and a pass-band at \(\omega_i\). A key characteristic of this high-pass filter is its reflectionless nature, ensuring a matched impedance of 50~\(\Omega\) across both the pass and stop bands. This property allows the first and third transmission line stages to be terminated in matched loads over the mTWPA's operational bandwidth.

\begin{figure}
\includegraphics[scale=0.7]{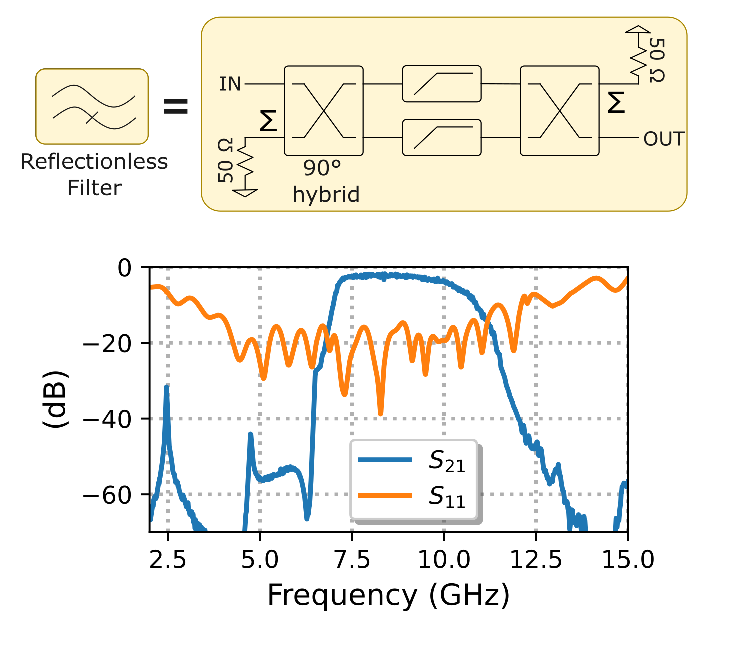}
\caption{\label{fig2}Balanced high-pass reflectionless filter implemented with two quadrature hybrids and two nominally identical multi-pole high-pass filters. Measurements of $S_{21}$ at room temperature show steep roll-off below $\omega_f/2\pi = 7.3$ GHz and $S_{11}$ shows good return loss of better than 15 dB over a bandwidth of 4-11 GHz which includes the stop- and pass-bands of the filters.}
\end{figure}

The balanced reflectionless filter topology is realized using two quadrature hybrids and low-temperature co-fired ceramic (LTCC) high-pass filters, as illustrated in Figure \ref{fig2}. The quadrature hybrids are implemented using Lange couplers, which exhibit insertion losses of \(3 \pm 0.5\)~dB and a phase shift of \(91^\circ \pm 1^\circ\) between the output ports over a frequency range of 4--11~GHz. Two balanced LTCC multi-pole high-pass LC filters are connected between the output ports of the quadrature hybrids.

The balanced high-pass filters generate identical reflections in the stop-band, which combine in-phase at the isolation port of the hybrid. This isolation port is terminated with a matched load, while at the input (IN) port, the reflections combine out of phase, thereby minimizing return loss in the stop-band. In the pass-band, signals propagate through both high-pass filters with a \(90^\circ\) phase separation and enter the output ports of the second quadrature hybrid. At this stage, the in-phase components are constructively combined at the output (OUT) port, while the out-of-phase components are directed to the isolation port, which is terminated in a matched load.

Experimental measurements of the transmission (\(S_{21}\)) and reflection (\(S_{11}\)) parameters of the reflectionless filter at room temperature are presented in Figure \ref{fig2}. The reciprocal filter design ensures a well-matched impedance for the transmission lines of the first and third stages of the mTWPA which is essential to achieving high gain with minimal gain ripple.

\section{Experiment}
The mTWPA was implemented using two coplanar SQUID transmission line sections fabricated on separate silicon dies, which constituted the first and third stages of the mTWPA. The transmission line sections and the reflectionless filter were modularized and connected together allowing flexibility to reconfigure the mTWPA and characterize each component separately. Each transmission line section had a length of $l = 350a$. The designed circuit parameters for the transmission line sections were: $I_0 = 1.2\ \mu$A, $r = 6$, $C_0 = 45$ fF, and $C_{\rm gnd} = 110$ fF.

The critical current $I_0$ of the Josephson junctions was estimated using the Ambegaokar-Baratoff relation \cite{tinkham}, based on room-temperature normal-state resistance measurements from test junctions fabricated in the same process. The permittivity of the SiN$_x$ dielectric was characterized using test resonators to determine $C_{\rm gnd}$, while the Josephson junction capacitance $C_0$ was calculated based on a capacitance of 40 $\text{fF/}\mu\text{m}^2$. The mTWPA was characterized in a dilution refrigerator at a temperature of 20 mK. The experimental measurement setup is shown in Figure \ref{figA4} which enabled full S-parameter and noise characterization of the mTWPA.

\begin{figure}[t]
\includegraphics[scale=0.55]{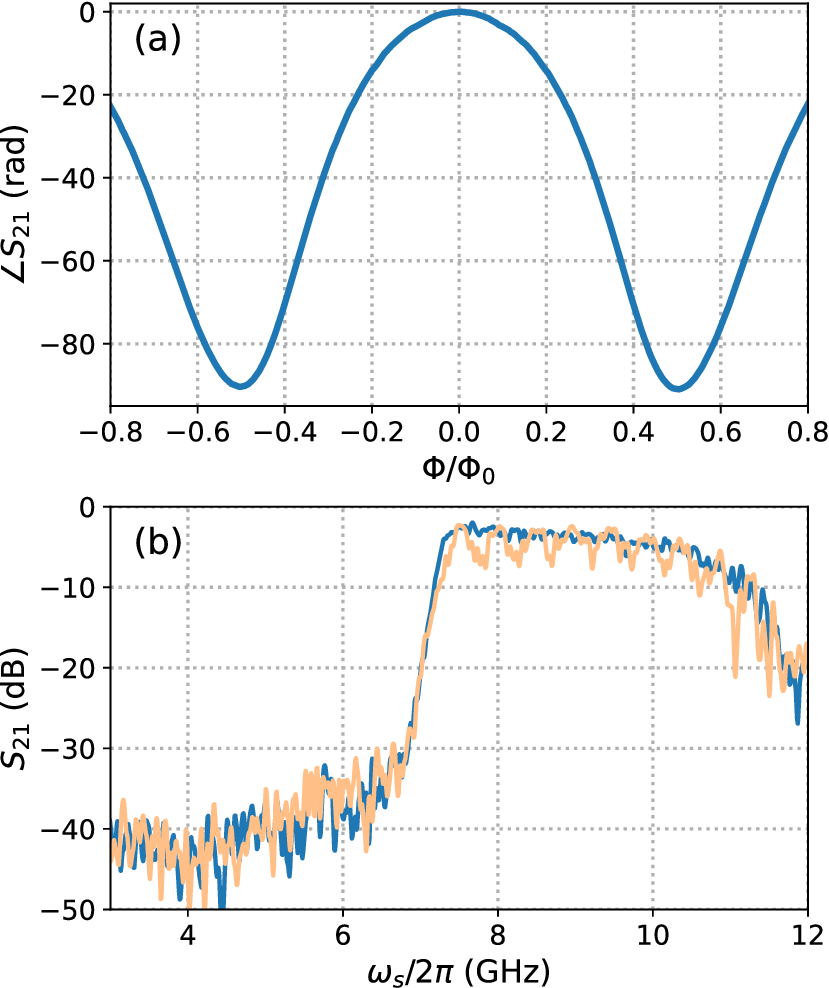}
\caption{\label{fig3}(a) Measured mTWPA Transmission $\angle S_{21}$ in radians versus $\Phi/\Phi_0$ at $\omega_s/2\pi = 7.4$ GHz. (b) Measured transmission $S_{21}$ of the mTWPA versus $\omega_s$, comparing flux bias values of $\Phi/\Phi_0 = 0$ (orange line) and 0.48 (blue line). Minimal ripple for $\Phi/\Phi_0 = 0.48$ indicate improved impedance matching.}
\end{figure}

The mTWPA was initially characterized via transmission measurements in the absence of an applied pump to determine the circuit’s electrical parameters. Transmission measurements were referenced against a thru-line, which could be switched in and out of the readout chain using cryogenic RF switches. An external superconducting solenoid applied a magnetic flux $\Phi$ to the SQUID loops of both transmission line stages. The measured phase response $\angle S_{21}$ as a function of $\Phi$ for the mTWPA is shown in Figure \ref{fig3}(a) at $\omega_s / 2\pi = 7.4$ GHz in the pass-band of the reflectionless filter. The SQUID loops of the first and third stage transmission line segments exhibited different flux couplings to $\Phi$. Based on independent measurements of each transmission line stage, as shown in Figure \ref{figA3}(a), the ratio of the flux periodicities between the first and third stages was determined to be $\Phi_{10} / \Phi_{30} = 1.26$. This discrepancy arose due to flux-focusing effects in the superconducting coplanar ground plane, where the lateral extension of the ground plane from the central conductor differed between the two stages \cite{fluxfocus1}.

The imbalance in flux coupling allowed for differential tuning of the first and third stages with an external applied magnetic flux. When the applied flux was set to $\Phi / \Phi_0 = 0.48$, the normalized flux biases of the SQUID loops for the first and third stages were $\Phi_1 / \Phi_{10} = 0.41$ and $\Phi_3 / \Phi_{30} = 0.52$, respectively. This configuration was designed to achieve a larger $|\gamma|$ in the third stage relative to the first stage. This adjustment was crucial because the same pump was used to drive both stages. Due to transmission losses in the first stage and the reflectionless filter, the pump power incident on the third stage was approximately 2 dB lower than that on the first stage. By tuning $\gamma$ through $\Phi$, it was possible to optimize the gain distribution across both stages.

Figure \ref{fig3}(b) presents $S_{21}$ of the mTWPA for $\Phi / \Phi_0 = 0$ (orange line) and $\Phi / \Phi_0 = 0.48$ (blue line). The high-pass response in transmission is attributed to the reflectionless filter. Just above $\omega_f$, the frequency where $\omega_p$ is positioned the insertion loss was measured to be 3 dB. The isolation level in the stop-band was limited by the noise floor of the VNA for this measurement. A standing wave pattern in the pass-band was observed when the mTWPA was tuned to $\Phi / \Phi_0 = 0$, which was attributed to impedance mismatches between the transmission line segments and the 50 $\Omega$ impedance of the reflectionless filter and the ports. Independent dispersion measurements of the first and third stages indicated an estimated characteristic impedance of 22 $\Omega$ which is consistent with the measured transmission ripple. For $\Phi / \Phi_0 = 0.48$, the standing wave ripple was minimized, indicating improved impedance matching of the transmission line segments to near 50 $\Omega$.

\begin{figure*}[t]
\includegraphics[scale=0.55]{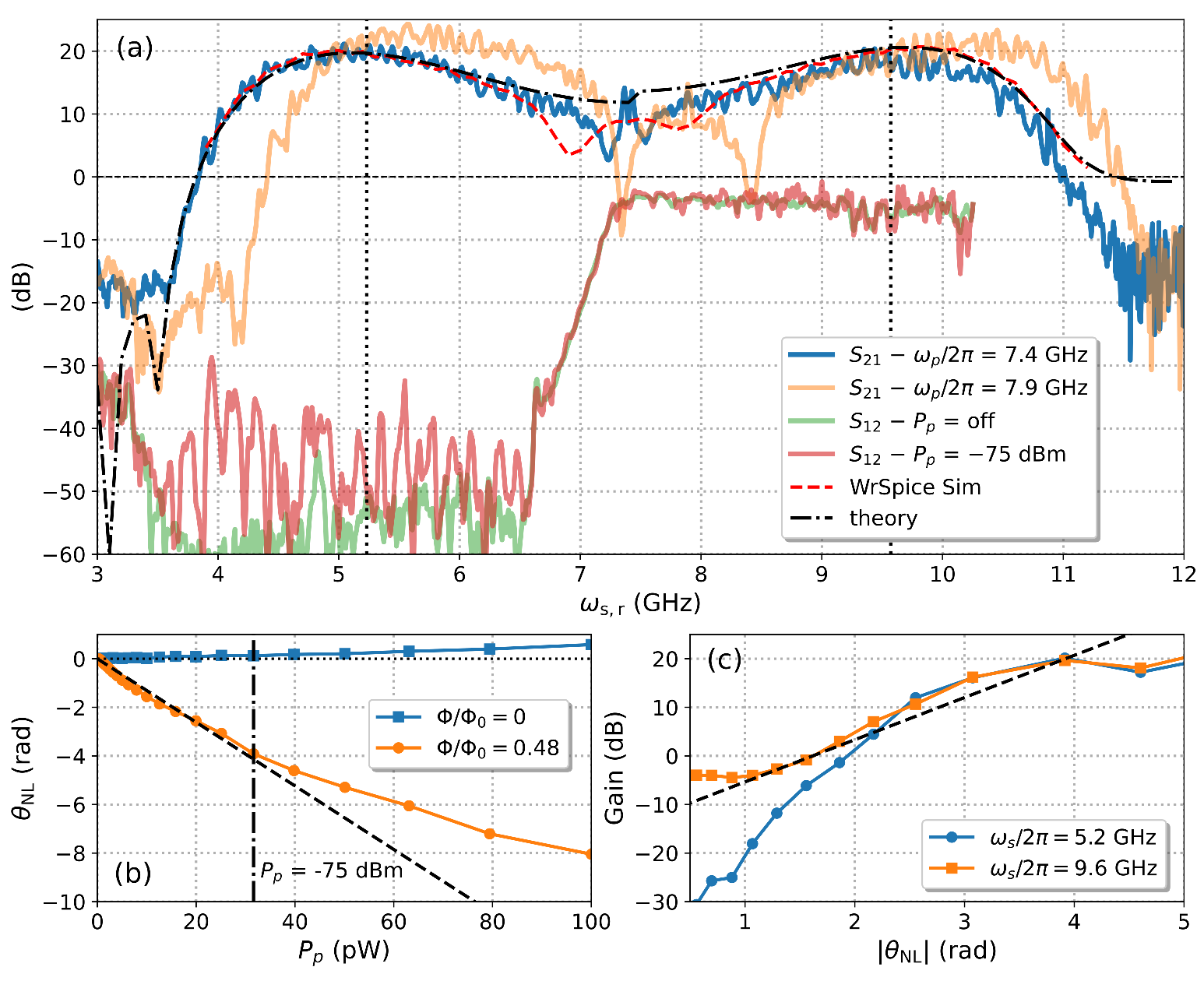}
\caption{\label{fig4}Measured nonreciprocal operation of the mTWPA. (a) Forward transmission $S_{21}$ versus $\omega_s$ for pump frequency ($P_p$) $\omega_p/2\pi = 7.4$ GHz (-75 dBm) (blue line) and 7.9 GHz (-74 dBm) (orange line). Vertical dashed lines indicated regions of forward 4WM parametric gain phase matching at $\omega_s/2\pi = 5.2$ and 9.5 GHz. The experimental data is in agreement with WRSpice circuit simulations (red dash line) and and theory Eqs. \ref{eq8} and \ref{eq11} (black dash-dot line). Reverse transmission $S_{12}$ through the mTWPA when the pump is on (red line) and off (green line). The reverse isolation is approximately 55 dB when the pump is off and 35 dB when the pump is on. The decrease in isolation with the pump signifies a potential higher order parametric process which is present. (b) Measured nonlinear phase $\theta_{\rm NL} = \angle S_{21}$ of $\omega_p$ versus $P_p$ for $\Phi/\Phi_0 = 0$ (blue squares) and 0.48 (orange circles). The difference in the two flux biases shows the tuning of the sign of the Kerr coefficient where a positive shift in $\theta_{\rm NL}$ is observed for $\Phi/\Phi_0 = 0$ and a negative shift for $\Phi/\Phi_0 = 0.48$. The dashed line is an expected dependence of $\theta_{\rm NL}(P_p)$, Eq. \ref{eq4}. Due to pump depletion effects at high powers there is a deviation from linear dependence above 30 pW. (c) Measured dependence of the forward signal gain versus $|\theta_{\rm NL}|$ for $\omega_s/2\pi = 5.2$ (blue circles) and 9.5 (orange squares) GHz. Dashed line illustrates the phased matched gain dependence $G_s \approx \exp (2\theta_{\rm NL})/16$ of the mTWPA.}
\end{figure*}

The key results of this work are the measured signal gain $S_{21} (\omega_s)$ of the mTWPA referenced to a thru-line, over a bandwidth which includes the pass-band and stop-band of the reflectionless filter of the mTWPA which is shown in Figure \ref{fig4}(a). The signal gain at $\Phi / \Phi_0 = 0.48$ and $\omega_p / 2\pi = 7.4$ GHz with pump power $P_p = -75 \pm 1$ dBm (blue line) is positioned just above $\omega_f$. The gain measurements demonstrate the broadband nature of the mTWPA, achieving a maximum gain of 20 dB over a bandwidth of 1.6 GHz. 

The inverse Kerr phase matching approach enables phase matching and maximum gain at frequencies $\omega_s/2\pi = 5.2$ GHz and 9.6 GHz, as indicated by the vertical dotted lines in Figure \ref{fig4}(a). This phase matching approach is necessary for achieving a broad operational bandwidth away from $\omega_f$ where $\omega_p$ is positioned slightly above. Despite the steep roll-off of 60 dB/GHz in the filter's stop-band the center of the gain lobe must be positioned at least 1.8 GHz from $\omega_f$ to ensure an adequate region of forward gain and reverse isolation simultaneously. An advantage of the inverse Kerr phase matching technique is that it does not rely on engineered dispersive features, thereby allowing for tunability of $\omega_p$, as demonstrated by an alternative pump tuning of $\omega_p / 2\pi = 7.9$ GHz and $P_p = -74 \pm 1$ dBm Figure \ref{fig4}(a) (orange line). The tunability of $\omega_p$ is limited by the need to position it near $\omega_f$ such that forward gain and reverse isolation can be achieved simultaneously with sufficient bandwidth. The absence of dispersive elements in the inverse Kerr phase matching approach, minimizes the adverse effects of impedance mismatches and suppresses gain ripples, which was observed to be less than 3 dB in experiment when biased with a suitable $P_p$. 

To determine the optimal $P_p$ for maximum signal gain, we measured the nonlinear phase shift $\theta_{\rm NL} = \angle S_{21} (P_p)$ of the pump in the mTWPA shown in \ref{fig4}(b) for flux bias $\Phi / \Phi_0 = 0$ (blue squares) and $\Phi / \Phi_0 = 0.48$ (orange circles). For $\Phi / \Phi_0 = 0$, the net $\theta_{\rm NL}$ is positive, similar to the behavior observed in a conventional TWPAs consisting of a chain of Josephson junctions. For $\Phi / \Phi_0 = 0.48$, $\theta_{\rm NL}$ acquires an inverse sign due to the sign change of $\gamma$ \cite{bellkerr}. The dependence of $\theta_{\rm NL}$ on $P_p$ is observed to be linear at low powers, in agreement with Eq. \ref{eq4}, where the dashed line represents the relation $\theta_{\rm NL} = \alpha_{\rm p1} (P_p) l + \alpha_{\rm p3} (P_p) l$, where $\alpha_{\rm p1}$ and $\alpha_{\rm p3}$ denote the SPM coefficients for the first and third stages, respectively. The disparity between $\alpha_{\rm p1}$ and $\alpha_{\rm p3}$ arises due to unequal $P_p$ delivered to and magnetic flux bias between the the first and third stages. For $P_p > 30$ pW, $\theta_{\rm NL}$ deviates from the expected linear behavior, which can be attributed to the generation of higher-order pump harmonics and pump depletion effects \cite{twpasim3, twpasim4_pumpharmonics}. The $P_p$ when $\theta_{\rm NL}$ deviates from linearity sets the upper limit on $P_p$ for optimal operation of the mTWPA.

\begin{figure}[t]
\includegraphics[scale=0.55]{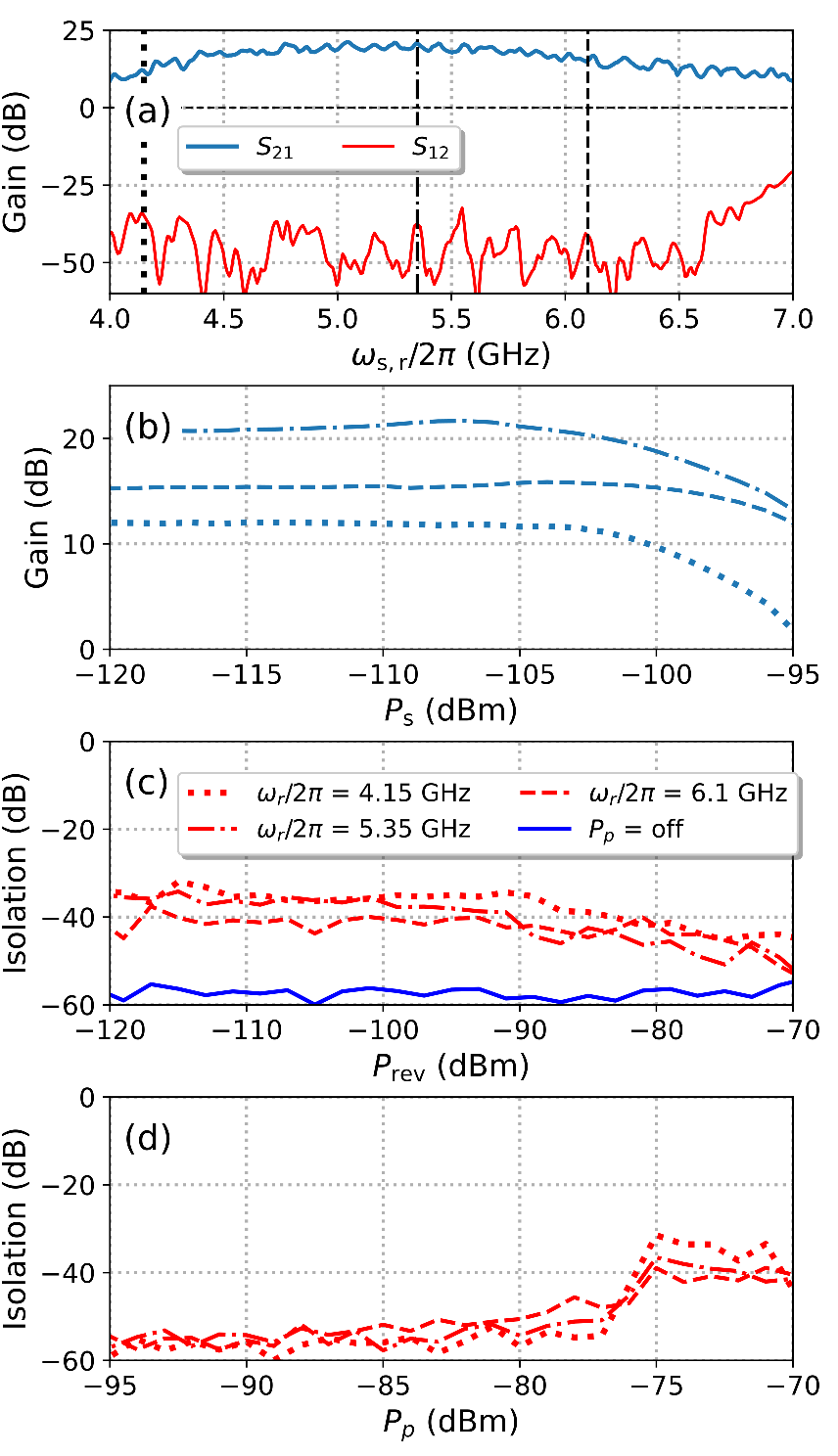}
\caption{\label{fig5}(a) mTWPA region of bandwidth where forward signal gain and reverse isolation can be achieved simultaneously. The vertical dash, dot, and dash-dot lines are markers and used as plot lines in subsequent panels in this figure to represent measurements taken at frequencies $\omega_s/2\pi = 4.15$, 5.35, and 6.1 GHz respectively. (b) Measurement of the gain $S_{21}$ versus forward signal power $P_s$. The saturation power is measured at the 1 dB compression point in gain which occurs at -101, -102, and -97 dBm for $\omega_s/2\pi = 4.15$, 5.35. and 6.1 GHz respectively. (c) Reverse isolation $S_{12}$ versus reverse signal power $P_{\rm rev}$. Reverse isolation when the pump is off (solid blue line). (d) Reverse isolation versus forward pump power $P_p$. At optimal $P_p = -75$ dBm for forward signal gain reverse isolation decreases to 35 dB.}
\end{figure}

The signal gain was experimentally measured as a function of $|\theta_{\rm NL}|$ and is shown in Figure \ref{fig4}(c) for $\omega_s / 2\pi = 5.2$ GHz (blue circles) and 9.6 GHz (orange squares). For \( \omega_s / 2\pi = 9.6 \) GHz and \( |\theta_{NL}| < 1 \) radian, where phase matching does not occur, the signal gain is negligible, and \( S_{21} \) is dominated by transmission losses. In the phase-matched regime where \( |\theta_{NL}| > 1 \) radian, the signal gain follows an exponential dependence $G_s \approx \exp(2|\theta_{\rm NL}|)/16$ as indicated by the dashed line. For $\omega_s / 2\pi = 5.2$ GHz, when $|\theta_{\rm NL}| < 2$ radian, phase matching is not achieved, and the signal gain remains negligible due to the isolation of the signal by the reflectionless filter. In the phase-matched regime, where $|\theta_{NL}| > 2$ radians, the signal experiences gain and follows the expected exponential dependence. At $|\theta_{\rm NL}| > 4$ radians, pump depletion effects become significant, causing the signal gain to saturate and an in increase in gain ripple and mTWPA noise was observed in experiment. 

For $\omega_s < \omega_f$, which lies within the stop-band of the reflectionless filter, the original forward-traveling signal does not propagate through the mTWPA as it is blocked by the reflectionless filter. Instead, within the first stage of the mTWPA, the signal generates an idler mode that resides in the pass-band of the reflectionless filter and is thus allowed to propagate with the pump to the third stage. In the third stage, the idler interacts with the pump to regenerate a signal at the output, thereby recovering an amplified signal with a transfer function given by Eq. \ref{eq11}. The measured signal gain for $\omega_p / 2\pi = 7.4$ GHz shown in Figure \ref{fig4}(a) is in good agreement with Eq. \ref{eq11} (black dash-dot line) for $\omega_s < \omega_f$. For $\omega_s > \omega_f$, the signal lies within the pass-band of the reflectionless filter, the forward-propagating wave is amplified in accordance with the standard gain relation described by Eq. \ref{eq8} and is in agreement with experiment (black dash-dot line) for $\omega_s > \omega_f$. The measured gain of the mTWPA is in good agreement with WRSpice simulations (Appendix \ref{AppendixE}) of the mTWPA circuit shown in Figure \ref{fig4}(a) (red dashed line).

The reverse isolation $S_{12}(\omega_r)$ of the mTWPA is presented in Figure \ref{fig4}(a) for both the pump on (red line) and pump off (green line) conditions. For $\omega_r > \omega_f$, corresponding to the pass-band of the reflectionless filter, the insertion loss is measured to be 3 dB. This loss accounts for the combined contributions of the filter and the transmission line losses of the first and third stages of the mTWPA. For $\omega_r < \omega_f$, which falls within the stop-band of the reflectionless filter, the isolation exceeds 35 dB across the mTWPA's operational bandwidth when the pump is on and improves to better than 50 dB when the pump is off. The observed reduction in isolation when the pump is on is likely attributed to non-phase matched parametric processes which are not completely suppressed in the mTWPA.

For $\omega_s < \omega_f$ the forward transmission $S_{21}(\omega_s)$ exhibits a nearly identical response to $S_{12}(\omega_r)$ in the pump-off state, as the mTWPA remains reciprocal under these conditions at low signal powers. The signal extinction ratio, defined as the ratio between the maximum gain when the pump is on and the maximum forward isolation when the pump is off, is measured to be 72 dB at $\omega_s / 2\pi = 5.2$ GHz. 

The gain and isolation characteristics of the mTWPA were measured against various signal powers, at $\Phi / \Phi_0 = 0.48$, $\omega_p / 2\pi = 7.4\, \text{GHz}$, and $P_p = -75\, \text{dBm}$ which are shown in Figure \ref{fig5}. Figure \ref{fig5}(a) The vertical lines indicate three frequencies $\omega_{\rm s,r} = 4.15$ (dotted), 5.35 (dash-dot), and 6.1 GHz (dash) where measurements of the signal gain and isolation of the mTWPA were performed. 

The forward signal gain of the mTWPA versus signal power $P_s$ is shown in Figure \ref{fig5}(b). The dynamic range is characterized by the 1 dB compression point $P_{\rm 1 dB}$, defined as the input signal power at which the gain drops by 1 dB. This is given as -101, -102, and -97dBm for $\omega_s/2\pi$ = 4.15, 5.35, and 6.1 GHz respectively. The measured dynamic range of the mTWPA is comparable to demonstrated TWPA performance in literature \cite{Rochphotonic,Rochrevkerr,whitetwpa,twpascience}. 

The $S_{12}(\omega_r)$ data in Figure \ref{fig5}(a) shows a series of peaks that are above the isolation level when the mTWPA is turned off and the three $\omega_r$ are marked with vertical lines. The reverse isolation $S_{12}(\omega_r)$ versus the reverse signal power $P_{\rm rev}$ is shown in Figure \ref{fig5}(c). For a counter-propagating reverse signal relative to the forward pump, there is significant phase mismatch in the 4WM process minimizing the effects of reverse parametric processes. The isolation observed is primarily due to the transfer function of the reflectionless filter. Since the filter is a linear electrical circuit, the reverse isolation extends to high $P_\text{rev}$. The observed increase in isolation is attributed to increased losses in the first and third stages of the mTWPA, resulting from reverse signal currents approaching $rI_0$ of the large Josephson junctions in the SQUIDs.

The peaks in the $S_{12}(\omega_r)$ measurements shown in Figure \ref{fig5}(a) dependence on $P_p$ are presented in Figure \ref{fig5}(d). For $P_p \lesssim -80$ dBm, the mTWPA is effectively off, exhibiting insignificant forward gain and minimal reverse parametric processes. When $P_p \gtrsim -77$ dBm, the isolation begins to decrease, dropping by 20 dB when $P_p$ reaches -75 dBm, which is the optimal tuning of the mTWPA. This reduction in isolation might be attributed to inefficient parametric processes becoming active at the operating point of the mTWPA. A detailed analysis of these parametric processes which contribute to the decrease in the reverse isolation of the mTWPA is beyond the scope of this work.

\begin{figure}[t]
\includegraphics[scale=0.6]{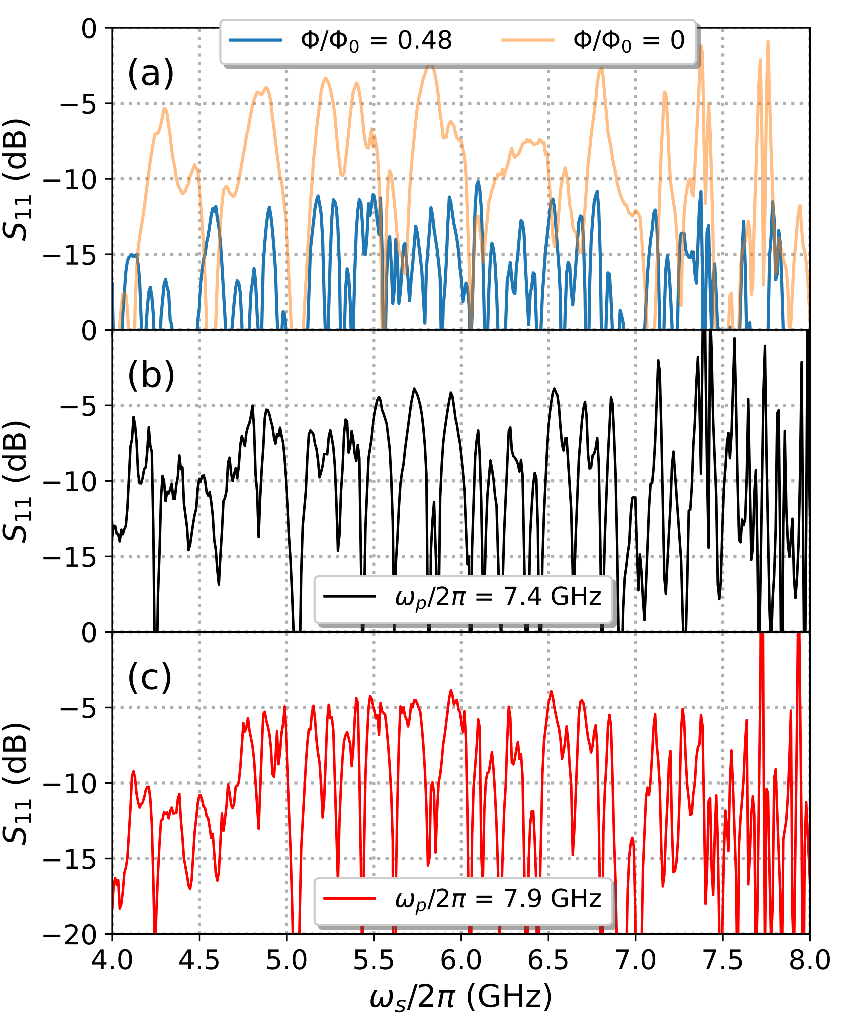}
\caption{\label{fig6} mTWPA return loss $S_{11}$ measurements (a) $S_{11}$ of the mTWPA versus $\omega_s$ for pump off and magnetic flux biasings of $\Phi/\Phi_0=0$ (orange line) and $\Phi/\Phi_0=0.48$ (blue line) representing impedance mismatch and matching conditions respectively. (b) and (c) $S_{11}$ of mTWPA with pump on when $\omega_p/2\pi~(P_p) = 7.4$ GHz (-75 dBm) and 7.9 GHz (-74 dBm) respectively. Measured increase in return loss ($\sim$6 dB) with pump on signifies the presence of an amplified backward traveling wave due to reflection at the first stage reflectionless filter interface.}
\end{figure}

Return loss measurements $S_{11}$ of the mTWPA in the absence of the pump are shown in Figure \ref{fig6}(a) for $\Phi / \Phi_0 = 0$ (orange line) and $\Phi / \Phi_0 = 0.48$ (blue line). When $\Phi / \Phi_0 = 0$, the return loss is measured to be better than 4 dB, which is in agreement with the first stage impedance of 22 $\Omega$ estimated from dispersion measurements. At $\Phi / \Phi_0 = 0.48$, the return loss improves to better than 12 dB, indicating a closer match to 50 $\Omega$ at the input port. From dispersion measurements an impedance of the first stage was estimated to be 46 $\Omega$ at $\Phi / \Phi_0 = 0.48$. The application of a pump is expected to cause a decrease in the impedance of the first-stage transmission line due to the inverse Kerr dependence on $P_p$. 

In practical TWPA applications, unavoidable impedance mismatches exist between the TWPA transmission line and the ports. This mismatch causes a portion of the amplified forward propagating signal to be reflected at the output port and propogate backward through the TWPA. This backward-propagating signal can be equal to or exceed the level of the input signal to the TWPA \cite{Rochtwpaiso}.  Since the backward-traveling waves have the same frequency as the forward, conventional filtering methods are ineffective. The most practical approach to mitigating the impact of backward traveling waves is the use of external isolators at the TWPA input.

In the mTWPA, small but nonzero reflections occur at the interface between the first-stage and the reflectionless filter causing a fraction of the amplified forward-propagating wave to reflect back toward the input port. Since the first stage of the mTWPA contributes only 10 dB of the total 20 dB signal gain, the return loss is improved compared to a conventional TWPA with the same 20 dB gain, transmission line loss, and output port return loss. 

Measurements of $S_{11}(\omega_s)$ at $\omega_p / 2\pi = 7.4$ GHz and $P_p = -75$ dBm indicate an increase in return loss to better than 5 dB as shown in Figure \ref{fig6}(b), when compared to the pump-off condition. A second pump tuning of $\omega_p / 2\pi = 7.9$ GHz and $P_p = -74$ dBm exhibit similar performance, as shown in Figure \ref{fig6}(c). The measured return loss is comparable to other approaches aiming to realize TWPAs with intrinsic reverse isolation via parametric up-conversion processes \cite{Rochtwpaiso,aumentadotwpaiso}. The improved return loss in these TWPAs is aided by the high transmission line losses reducing the backward propagating signal. The mTWPA exhibits a first-stage transmission line loss of 0.5 dB at 5 GHz, Figure \ref{figA3}(b) which does very little to isolate reverse propagation. As TWPA transmission line losses improve, return losses will get worse for devices without reverse isolation characteristics and the need for isolation will increase. In Appendix \ref{AppendixA}, we discuss simple design improvements that could significantly enhance the return loss of the mTWPA presented here through redistribution of the signal gain between the two stages.

\begin{figure}[t]
\includegraphics[scale=0.55]{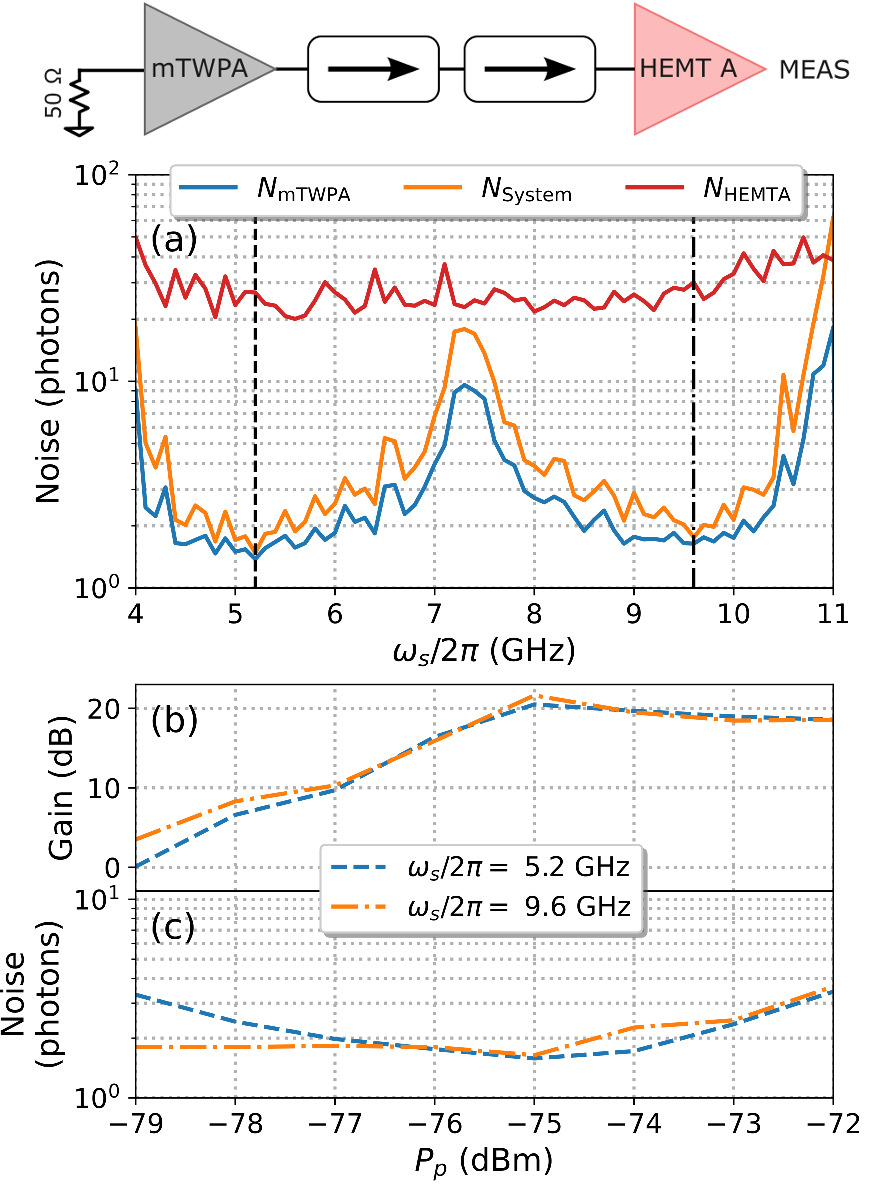}
\caption{\label{fig7} Noise measurement of mTWPA performed with a noise calibrated HEMT A amplifier with isolators shown at top of plot. (a) Measured noise in photon units of the mTWPA (blue line), system noise (orange line) and effective HEMT A noise (red line). Vertical dash and dash-dot lines indicate regions of phase matching above and below $\omega_f$. Measured mTWPA gain (b) and mTWPA noise (c) versus $P_p$ at $\omega_s/2\pi = 5.2$ and 9.6 GHz.}
\end{figure}

\section{Noise Measurements}

The noise performance of the mTWPA was evaluated using an amplifier chain characterized through a Y-factor measurement \cite{pozar}, employing a broadband thermal noise source tunable between 150 mK and 4 K. This measurement determined the effective noise temperature $T_{\rm HEMTA}$ of the amplifier chain, which is primarily dominated by the noise of the HEMT amplifier A. The measured $T_{\rm HEMTA}$ of the readout amplifier chain in photon units $N_{\rm HEMTA}=k_BT_{\rm HEMTA}/\hbar\omega_s$ is shown in Figure \ref{fig7}(a) (red line).

The noise contribution of the mTWPA was determined using the signal-to-noise ratio improvement method \cite{whitetwpa}, wherein both the signal power gain $G_s$ and noise gain $G_{\rm noise}$ at the output of the readout chain were measured as the mTWPA was switched on and off. The system and mTWPA noise were determined using the relations $T_{\rm sys} = G_{\rm noise} T_{\rm HEMTA}/{G_s}$ and $T_{\rm TWPA} = T_{\rm HEMTA} (G_{\rm noise} - 1)/{G_s}$ and are presented in Figure \ref{fig7}(a) in photon units as the orange and blue lines, respectively. Over the 4.5–6 GHz bandwidth of the mTWPA, the noise remains below two photons, near the one photon standard quantum limit. Given that the gain of the mTWPA within this bandwidth is approximately 20 dB, the system noise is about two photons.

Demonstrating a system noise of approximately twice the quantum limit in an amplifier chain with a relatively high effective noise of $N_{\rm HEMTA}\approx 25$ photons highlights the versatility of the mTWPA in achieving an improved signal-to-noise ratio, even in a non-optimized readout chain. Additionally, in the upper gain lobe where the mTWPA does not provide reverse isolation the noise remains below two photons. The measured noise performance of the mTWPA, despite the added loss and complexity of the reflectionless filter, demonstrates an intrinsic noise level comparable to some of the lowest reported for single-stage TWPAs \cite{whitetwpa,twpascience,Rochrevkerr}.

The measured signal gain and mTWPA noise in the phase matching region, indicated by the vertical dashed ($\omega_s/2\pi = 5.2$ GHz) and dash-dot ($\omega_s/2\pi = 9.6$ GHz) lines in Figure \ref{fig7}(a) are shown versus $P_p$ in Figure \ref{fig7}(b) and (c). When the mTWPA operates at $P_p = -75$ dBm, the noise is the lowest at 1.7 photons. As $P_p$ increases, so does the noise, until pump saturation due to high harmonic generation, and parametric coupling to high frequency products. When $P_p < -77$ dBm, the noise at $\omega_s/2\pi = 9.6$ GHz remains constant as both $G_s$ and $G_{\rm noise}$ decrease proportionally as is observed in conventional single stage TWPAs. However, at $\omega_s/2\pi = 5.2$ GHz, the noise increases to 3.2 photons at $P_p = -79$ dBm, coinciding with a drop in $G_s$ to 0 dB. Most of the noise observed at the mTWPA output at $\omega_s/2\pi = 5.2$ GHz arises from idler-to-signal conversion in the third stage (Eq. \ref{eq10}) of amplified noise from the first stage of the mTWPA. The quantum noise at the input to the third stage in the $\omega_s/2\pi = 5.2$ GHz band makes a small contribution. The measured output noise was referred back to the input through the experimentally determined $G_s$ of the mTWPA. For $P_p < -77$ dBm, $G_s (\omega_s < \omega_f) < G_s (\omega_s > \omega_f)$ as shown in Figure \ref{fig7}(b), this results in a relative increase in input referred noise at frequencies below $\omega_f$.

\begin{figure}[t]
\includegraphics[scale=0.55]{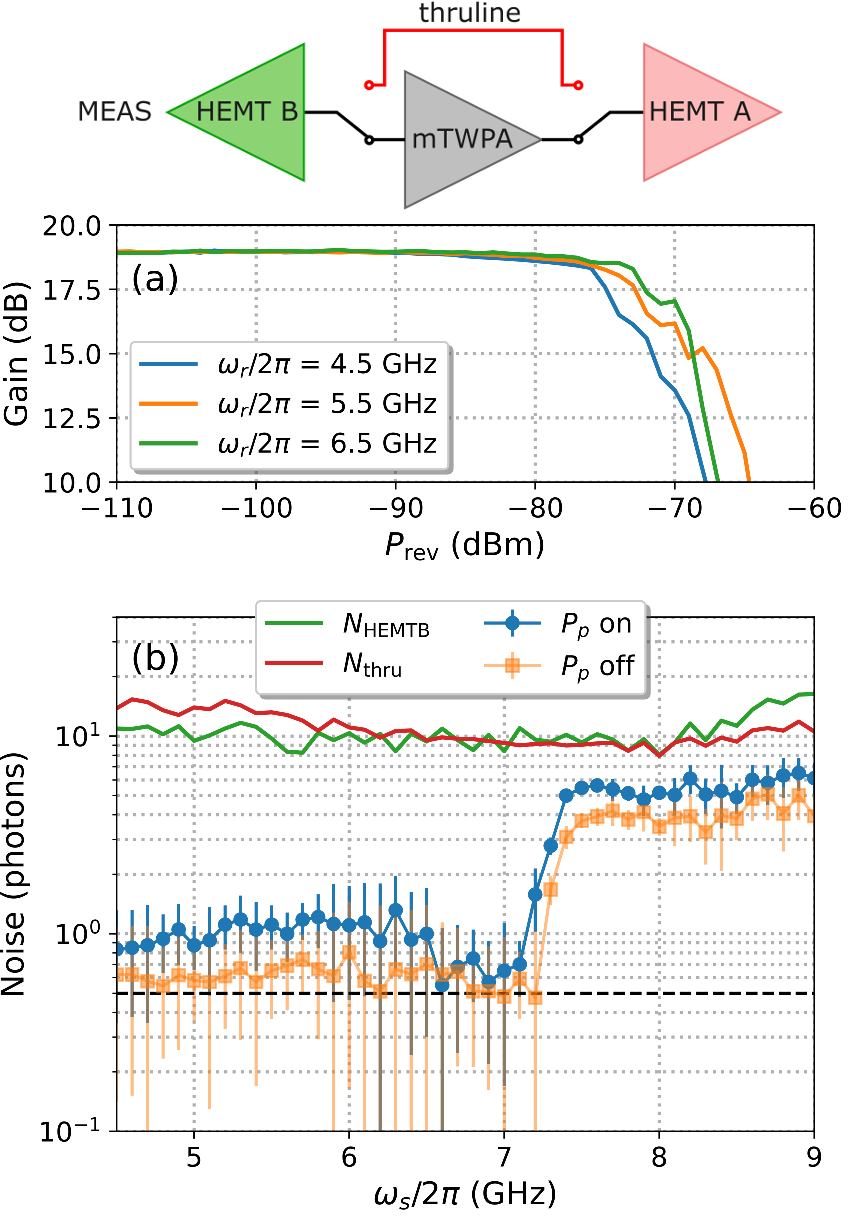}
\caption{\label{fig8} Noise and gain measurement of mTWPA performed without isolation between mTWPA and HEMT A. HEMT B used to directly measure input noise of mTWPA from backward-propagating noise from HEMT A and mTWPA. (a) Forward signal gain at $\omega_s/2\pi = 5.2$ GHz for $\omega_p/2\pi = 7.4$ GHz and $P_p = -75$ dBm versus $P_{\rm rev}$ at $\omega_r/2\pi = 4.5$, 5.5, and 6.5 GHz. Signal gain is resilient for $P_{\rm rev}$ up to -76 dBm where the gain drops. (b) Measured noise in photon units at reference plane A for the thru line between HEMT A and HEMT B (red line), the mTWPA with pump on (blue circles), and mTWPA with pump off (orange squares). The noise of HEMT B, characterized with a Y-factor measurement is shown as a green line}
\end{figure}

An ideal TWPA with reverse isolation should function as a near quantum-limited amplifier, positioned between the HEMT amplifier and the quantum device under measurement, without requiring isolators before or after the TWPA. The TWPA must be able to tolerate the total broadband noise from the HEMT without disturbing the in-band forward signal gain while sufficiently isolating this noise from reaching the device under measurement.

Figure \ref{fig8}(a) presents the forward signal gain at $\omega_s/2\pi = 5.2$ GHz, measured for reverse propagating signals at $\omega_r/2\pi = 4.5$ GHz (blue line), 5.5 GHz (orange line), and 6.5 GHz (green line) versus $P_{\rm rev}$. The reverse propagating signal only affects the parametric gain of the third stage of the mTWPA, since the signal lies in the stop band of the reflectionless filter, preventing interference with the first stage. The forward signal gain is suppressed by 1 dB at $P_{\rm rev} = -76$ dBm. This power level is significantly higher than the noise power emitted by typical HEMT amplifiers used in circuit quantum electrodynamics (cQED) measurement setups. This shows that the mTWPA can effectively operate without isolation between the mTWPA and the HEMT.

The noise at the input of the mTWPA was measured without isolators in the readout chain as shown in Figure \ref{fig8}, utilizing the measurement setup shown in Figure \ref{figA4}. The noise at the input of the mTWPA was measured directly with the HEMT B readout chain. The effective noise temperature of HEMT B $T_{\rm HEMTB}$, was characterized with a Y-factor measurement using a broadband noise source, which was switched into the readout chain. The effective noise of HEMT B is shown in Figure \ref{fig8} (b) (green line). 

The first input noise measurement of the mTWPA involved inserting a thru-line between HEMT A and HEMT B to quantify the noise emanating from HEMT A, Figure \ref{fig8}. The measured noise gain $G_{\rm noise}$, was determined at the input reference plane A (Figure \ref{figA4}) to the mTWPA by measurement of the noise power when the thru-line was switched in compared to the measured noise power of a cold load. The noise at reference plane A of the thru-line was approximately 12 photons $N_{\rm thru} = (G_{\rm noise} - 1) N_{\rm HEMTB}$, as shown in Figure \ref{fig8}(b) (red line). The effective noise of HEMT A, measured as $N_{\rm HEMTA}\approx$ 25 photons in Figure \ref{fig7}(a), includes cable and isolator losses between HEMT A and the mTWPA output reference plane B (Figure \ref{figA4}). Accounting for these losses along with the thru-line loss of $\sim 0.2$ dB will reduce the noise from HEMT A to 12 photons at reference plane A in the measurements without isolators. This noise level aligns well with the measured $N_{\rm thru}$ and the manufacturer’s noise specifications for HEMT A \cite{lnfhemt1}.

The noise at the input of the mTWPA was measured with the mTWPA switched into the measurement circuit. The noise measured when the pump was switched on and off, are presented in Figure \ref{fig8}(b) as blue circles and orange squares, respectively, including error bars. For $\omega_s / 2\pi < 7.1$ GHz, the mTWPA exhibits sufficient isolation to suppress the noise from HEMT A. The measured noise at the input of the mTWPA is approximately half a photon when the pump is off and remains below 1.3 photons when the pump is on. 

The ideal noise at the input of the mTWPA is the quantum noise of one-half photon. This noise level aligns with the measured noise of the mTWPA when the pump is off. When the pump is on, a half-photon of quantum noise enters the first stage of the mTWPA at both $\omega_s$ and $\omega_i$ frequency bands. Through parametric gain at $\omega_s$ and idler-to-signal down-conversion, described by Eq. \ref{eq10}, the total noise at the output of the first stage of the mTWPA is additive and reflects off a non-zero impedance mismatch of the first stage - reflectionless filter interface. The reflected amplified noise propagates backward through the first stage and exits port 1 of the mTWPA, and contributes to the total measured noise at the input. Given the first stage signal gain of 10 dB, an estimate of 15 dB return loss at the first stage - reflectionless interface, and the added noise of the first stage transmission line of the mTWPA, it is conceivable that the noise at the input of the mTWPA can reach greater than one photon at its input in agreement with measurements in Figure \ref {fig8}(b). WRSpice simulations of the input noise of the mTWPA are shown in Figure \ref{figA1} which estimates a first stage input noise of $0.8$ photons for $l=350a$. These simulations do not take into account added noise due to transmission line losses which was observed in the experiment of the mTWPA.

For $\omega_s / 2\pi > 7.4$ GHz, the noise measured at the input of the mTWPA, both with the pump on and off, predominantly originates from HEMT A, attenuated by losses in the mTWPA, totaling approximately 3 dB. This attenuation reduces the HEMT A noise measured by $N_{\rm thru}\approx$ 10 photons to 4.5 photons when the pump is off. When the pump is on, the noise level increases by 1 photon, which is the addiditive noise contribution from the mTWPA. In this frequency region, the noise behavior of the mTWPA is analogous to that of a conventional TWPA with minimal transmission line loss which attenuates noise from the warmer HEMT. However, relying solely on transmission line losses in conventional TWPAs is insufficient to suppress HEMT amplifier noise, necessitating additional in-band isolation at the input and output of conventional TWPAs. For the mTWPA presented here, the highest noise level at maximum gain in the phase-matching region is 1.3 photons. This performance can be further improved by redistributing the gain between the stages and improving impedance match between the first stage and reflectionless filter of the mTWPA, as discussed in Appendix \ref{AppendixA}.

\section{Discussion}
We have presented a mTWPA design that exhibits passive in-band reverse isolation and has been experimentally demonstrated to achieve 20 dB of forward signal gain over 1.6 GHz bandwidth. The in-band isolation of backward-propagating signals was measured to be better than 35 dB. The mTWPA was designed with a three-stage architecture. The first and third stages consisted of coupled asymmetric SQUID nonlinear transmission lines employing the inverse Kerr phase-matching technique for signal-to-idler and idler-to-signal mode conversion respectively. The second stage incorporates a reflectionless high-pass filter that blocks both forward and reverse signal propagation in the stop-band while allowing the idler and pump in the pass-band to propagate to the third stage.  Phase matching a co-propagating signal and pump in a four-wave mixing process is crucial for these conversion processes and for achieving signal gain. In contrast, a reverse-propagating signal experiences a large phase mismatch with the forward propagating pump, minimizing parametric processes, and is isolated with a passive reflectionless filter.

Noise measurements demonstrate that the mTWPA achieves near-quantum-limited performance, with a noise level of 1.7 times the quantum limit. These results confirm that despite the increased complexity of the mTWPA design, near-quantum-limited noise performance can still be achieved. The demonstrated level of isolation effectively suppresses noise from the warmer HEMT amplifier stage. The mTWPA exhibits reduced in-band amplifier noise that propagates backward through the mTWPA that would contribute to back-action on sensitive quantum devices under measurement. 

Design improvements to the mTWPA suggest that implementing the first stage as a buffer stage with near-unity gain would further reduce noise at the input of the mTWPA to a level approaching one-half a photon while also improving return loss. Such improvements would enable an mTWPA that can operate in a conventional measurement setup without requiring bulky isolators at its input or output, enhancing measurement quantum efficiencies and facilitating scalability of prototype quantum computer readout.

\begin{acknowledgments}
This work was supported in part by NSF grants DMR-1838979 and OSI-2328774. Device characterization was performed at the University of Massachusetts Boston Quantum Core facility, which is funded through a grant from the Massachusetts Technology Collaborative matching grant program. M.T. Bell acknowledges funding from Google, through a Google Academic Research Award.
\end{acknowledgments}

\appendix
\counterwithin{figure}{section}

\section{\label{AppendixA}mTWPA Design Optimization}

The mTWPA exhibited a return loss of 5 dB, as illustrated in Figure \ref{fig6}(b,c), and an excessive input noise level of 1.3 photons, as shown in Figure \ref{fig8}(b). In conventional TWPAs, return loss and input noise levels tend to be even greater due to the cumulative effects of total signal gain and output port reflections, which can lead to reflected signals exceeding the magnitude of the input signal. Both the suboptimal return loss and elevated noise level at the input to the mTWPA are primarily attributed to the signal gain in the first stage and the impedance mismatch between the first stage and the reflectionless filter interface. 

 Numerical simulations using WRspice \cite{wrspce_pub,wrspice,twpasim1,twpasim2,twpasim3} are presented to evaluate an optimized mTWPA design aimed at improving return loss and reducing noise at the input. To reduce return loss and minimize input noise of the mTWPA, the distribution of signal gain can be optimized between the first and third stages. By setting the signal gain in the first stage close to unity, this stage effectively functions as an input buffer while still enabling sufficient idler mode generation. In these numerical simulations, we constrain our analysis to the same electrical circuit parameters and flux biasing conditions as those used in the first and third stages of the experimentally demonstrated mTWPA. Enhancements in return loss and input noise are achieved by optimizing the $l$ of the first stage and the $Z_f$ of the reflectionless filter.

\begin{figure}[t]
\includegraphics[scale=0.55]{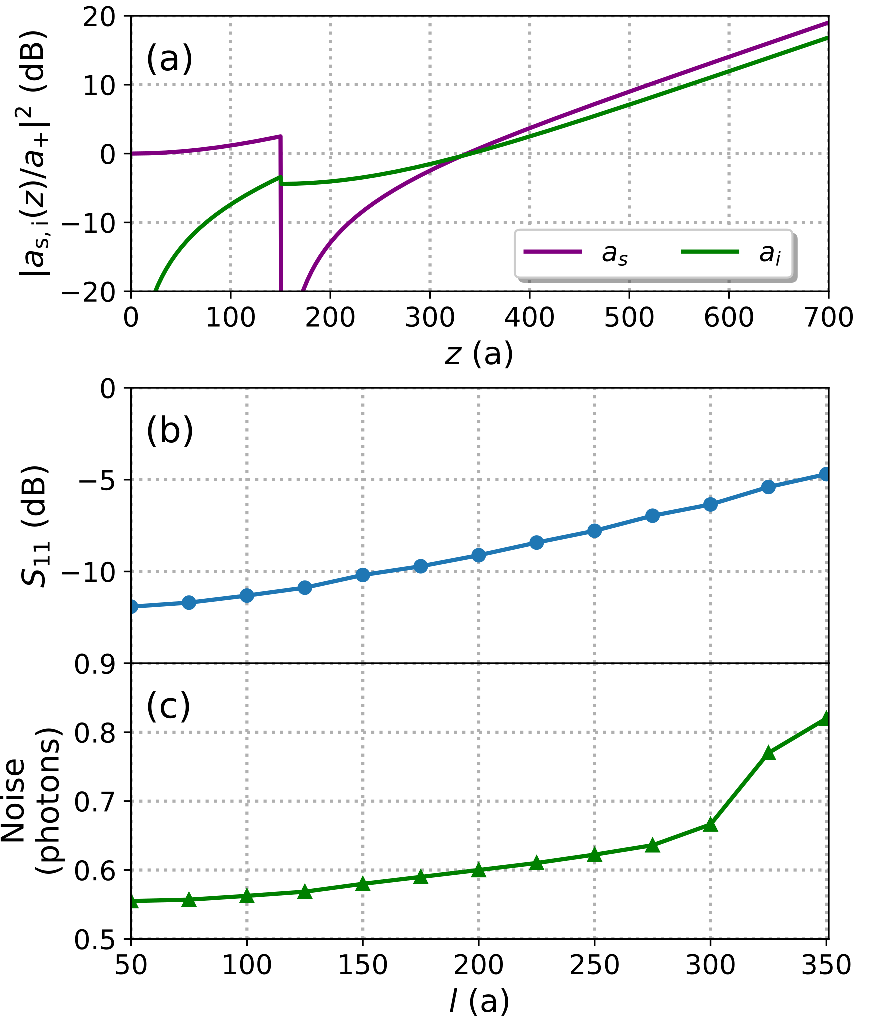}
\caption{\label{figA1} mTWPA circuit design optimization. (a) Power gains of signal and idler $|a_{\rm s,i}/a_+|^2$ relative to input signal versus position $z$ along mTWPA calculated from eq. \ref{eq8}, \ref{eq9}, and \ref{eq10} for first and third stage $l=150a$ and $550a$ respectively. (b) $S_{11}$ numerically simulated with WRspice versus first stage transmission line length $l$. (c) Noise in units photons at the input to the mTWPA simulated with WRspice versus first stage $l$.}
\end{figure}

Figure \ref{figA1}(a) presents estimates of the forward signal gain (purple line), as described by Eqs. \ref{eq9} and \ref{eq10}, for an mTWPA design with first and third stage transmission line lengths of $l = 150a$ and $l = 550a$, respectively. The calculated idler mode is depicted by the green line. By reducing the length of the first stage, the signal gain is limited to less than 2 dB, while in the third stage, the signal is effectively recovered from the idler mode, achieving an output signal gain of 20 dB.

With a reduction in $l$ of the first-stage, both the return loss and the noise at the input of the mTWPA are expected to decrease. WRspice simulations of $S_{11}$ are shown in Figure \ref{figA1}(b), which illustrate a reduction in return loss with decreasing $l$ of the first stage of the mTWPA. WRspice simulations of the noise at the input to the mTWPA are depicted in Figure \ref{figA1}(c) which show a decrease in noise with $l$. WRspice simulations show that for a practical first-stage length of $l = 150a$, the return loss is reduced to 10 dB and the noise at the input is reduced to 0.59 photons.

Further reduction in return loss and input noise can be achieved by optimizing the input impedance $Z_f$ of the reflectionless filter relative to the characteristic impedance $Z_0$ of the first stage of the mTWPA. For a flux bias of $\Phi_1 / \Phi_{10} = 0.41$, dispersion measurements of the first stage indicate that the characteristic impedance is $Z_0$ = 46 $\Omega$ at low $P_p$. When $P_p = -75$ dBm, $Z_0$ decreases to 42 $\Omega$ due to the inverse Kerr effect. 

\begin{figure}[t]
\includegraphics[scale=0.55]{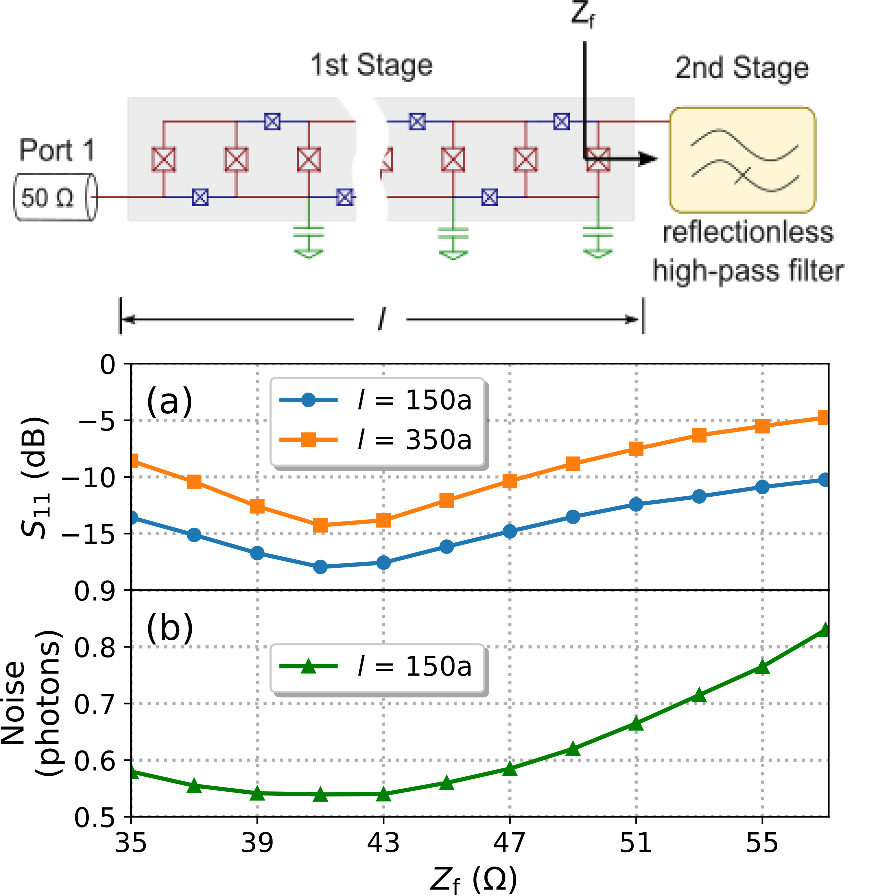}
\caption{\label{figA2} mTWPA circuit design optimization by improving impedance match between first stage and reflectionless filter with input impedance $Z_f$. (a) Reflection $S_{11}$ numerically simulated with WRspice for two first stage $l=$ 150$a$ (blue circles) and 350$a$ (orange squares) versus $Z_f$. (b) Noise in photon units numerically simulated with WRspice for first stage $l=150a$ versus $Z_f$.}
\end{figure}

WRspice simulations of $S_{11}$ for a first stage transmission line length of $l = 150a$ (blue circles) and $l = 350a$ (orange squares) are shown in Figure \ref{figA2}(a). A minimum return loss of 18 dB for $l = 150a$ is observed when $Z_f = 41~\Omega$, matching the impedance of the first stage transmission line of the experimentally measured mTWPA. WRspice simulations of the input noise for $l = 150a$ are presented in Figure \ref{figA2}(b). For $Z_f \approx 41~\Omega$, the input noise is minimized to 0.53 photons, approaching the quantum noise limit of a 50 $\Omega$ environment. Further improvements in return loss can be had with better matching of the first stage and reflectionless filter to the 50 $\Omega$ environment.

Improvements in the impedance match between the first stage transmission line of the mTWPA and the reflectionless filter can be predetermined and fixed during fabrication and do not depend on how the mTWPA is connected to an external circuit. This approach offers a significant advantage over conventional TWPA designs, where the output port match is contingent on the impedance of the external environment. In contrast, for the mTWPA, the output port match is less critical, as any in-band reflected signal on the output port is effectively isolated by the reflectionless filter and will not make it to the input port to degrade return loss or contribute to higher noise.

\section{\label{AppendixB}mTWPA Flux Dependence}
In the mTWPA characterized in this work, the transmission lines of the first and third stages were designed to be nominally identical. In the measured mTWPA there was a distinction between the two stages which was the extent to which the ground plane of the coplanar transmission line extended from the SQUID transmission line central conductor. This co-planar transmission line structural difference resulted in different periodicities in magnetic flux between the two stages due to different degrees of flux-focusing on the SQUID transmission line \cite{fluxfocus1}. 

\begin{figure}[t]
\includegraphics[scale=0.55]{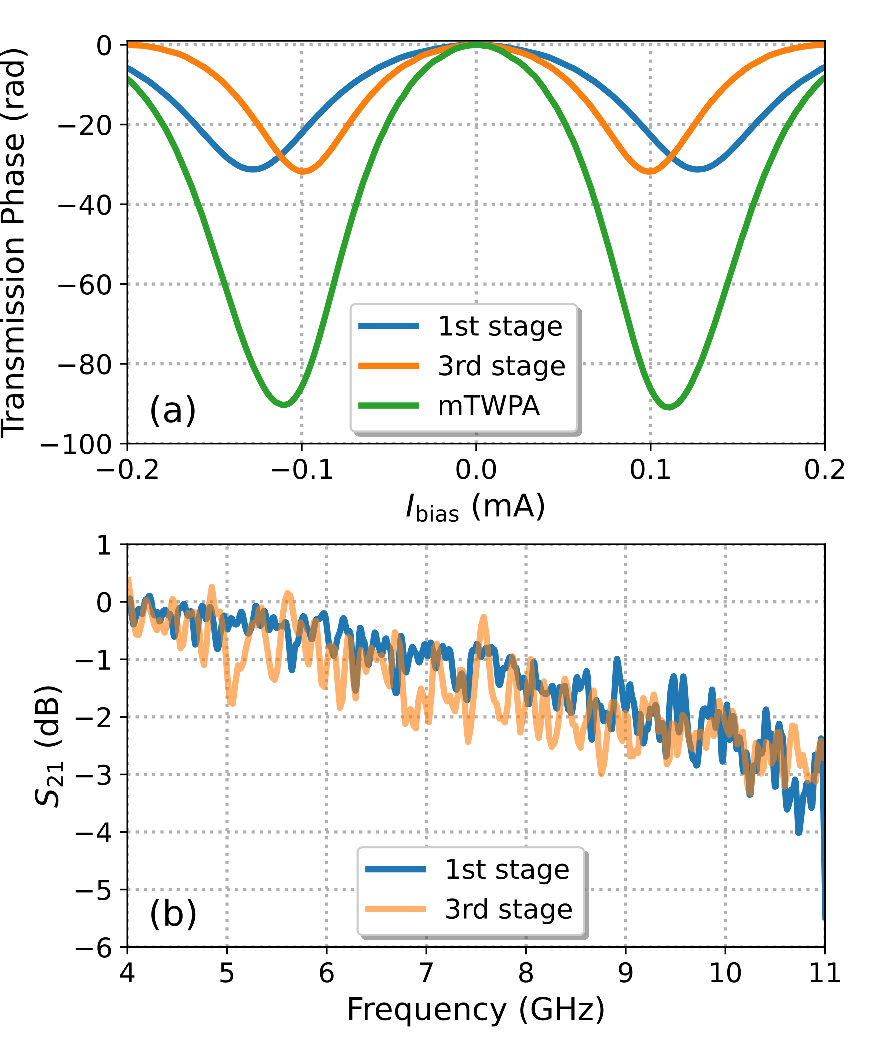}
\caption{\label{figA3} (a) Transmission phase versus DC current $I_{\rm bias}$ applied to external superconducting solenoid. $\angle S_{\Sigma 1}$ dependence for $\omega_s/2\pi = 5$ for the first stage (blue line) and $\angle S_{2\Sigma}$ for the third stage (orange line) measured independently. Overall $\angle S_{21}$ of the mTWPA measured at $\omega_s/2\pi = 7.4$. (b) Transmission $S_{21}$ of the first and third stages measured independently in a separate cooldown.}
\end{figure}

Figure \ref{figA3}(a) presents independent measurements of the first-stage $\angle S_{\Sigma 1}$ (blue line) and the third-stage $\angle S_{2\Sigma}$ (orange line) transmission phase as a function of the DC bias current $I_{\rm bias}$ applied to the external solenoid which flux-biases both stages simultaneously. The transmission measurements were performed at a signal frequency $\omega_s/2\pi = 5$ GHz, which lies within the stop-band of the reflectionless filter, thereby causing the signal to exit the $\Sigma$ port of the quadrature hybrids in Figure \ref{fig2}. The phase response measurements as a function of magnetic flux revealed periodicity in $I_{\rm bias}$ of $\Phi_{10} = 0.25$ mA and $\Phi_{30} = 0.2$ mA for the first and third stages, respectively. The nearly identical phase change at full frustration in both stages indicate comparable values of $k_s$ between the first and third stages, which is consistent with the nominally identical circuit parameters. Measurements of the overall transmission phase $\angle S_{21}$, obtained within the pass-band of the reflectionless filter at $\omega_s/2\pi = 7.4$ GHz, are represented by the green line, exhibiting a periodicity in magnetic flux of 0.22 mA.

Independent transmission measurements for the first (blue line) and third (orange line) stages are shown in Figure \ref{figA3}(b). These measurements were conducted on the individual stages in a separate cooldown before they were integrated into the mTWPA package. The observed insertion loss suggests a dielectric loss tangent slightly exceeding the expected value of $\tan\delta \approx 0.0005$ obtained from test resonator measurements. The overall low insertion loss contributes to the feasibility of driving both stages with a single pump and enhances the mTWPA's capability to achieve near-quantum-limited noise performance.

\section{\label{AppendixC}Measurement Setup}

\begin{figure}[t]
\includegraphics[scale=0.46]{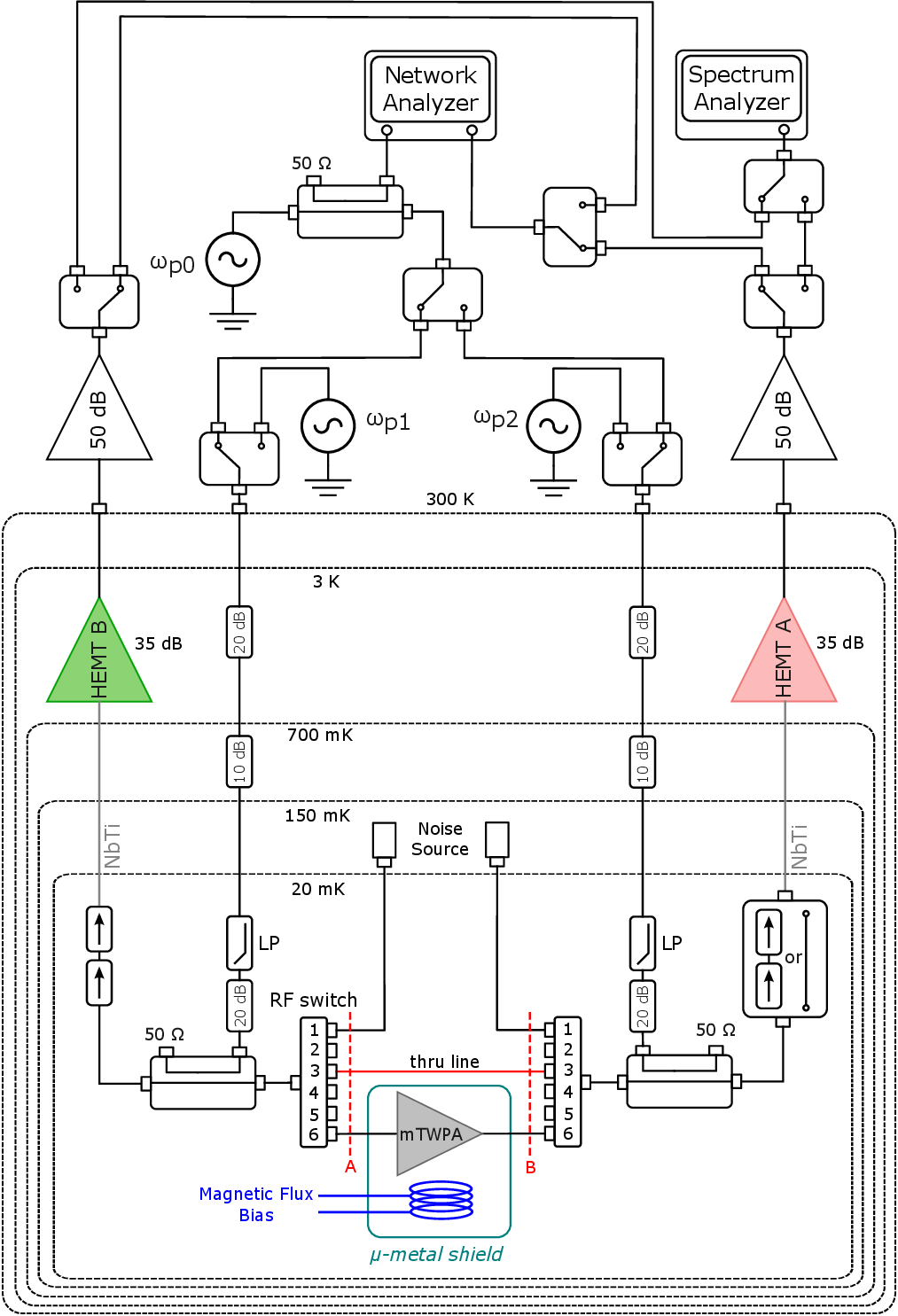}
\caption{\label{figA4} Experimental setup used to characterize the mTWPA. RF switches at room temperature utilized to switch between the various measurement and mTWPA pumping configurations. The experimental setup through latching RF switches allowed for different low temperature calibration and measurement configurations. Two separate cooldowns were performed to characterize the mTWPA with and without cryogenic isolators in the HEMT A readout chain.}
\end{figure}

The measurements were conducted in a Bluefors LD400 dilution refrigerator at a base temperature of 20 mK with the measurement setup shown in Figure \ref{figA4}. The mTWPA was flux-biased using an external superconducting solenoid. Both the mTWPA and the external solenoid were enclosed within a cryogenic $\mu$-metal magnetic shield to mitigate noise. The cryogenic measurement system was configured for two-port $S$-parameter characterization of the mTWPA. Signals originating at room temperature were transmitted to the mixing chamber via coaxial cables, with a standard configuration of attenuators to suppress noise. Return signals were amplified through two separate readout chains (HEMT A and HEMT B) before reaching room temperature. Each readout chain consisted of double-junction Low Noise Factory (LNF) isolators, NbTi superconducting coaxial cables, and LNF HEMT amplifier, followed by amplification via room-temperature amplifiers. In a separate cooldown designed to characterize the input noise of the mTWPA, the HEMT A readout chain was reconfigured to replace the double-junction isolator with a thru-line to mimic a measurement configuration without isolators.

To facilitate switching between different measurement conditions, two Radiall SP6T latching RF switches were employed at reference planes A and B (denoted by red dashed lines in Figure \ref{figA4}), which designate the measurement input and output reference planes of the mTWPA respectively. These switches allow for switching between the mTWPA, a thru-line, noise sources, and thru-line-reflect (TLR) microwave calibration references (not shown, but the line and short positioned at RF switch ports 2 and 3). The noise source for Y-factor measurements consisted of a Cu plate mounted on the 150 mK intermediate cold plate of the dilution refrigerator, with a weak thermal link. This plate was equipped with two Quantum Microwave QMC-CRYOTEM-DC18NM 50 $\Omega$ NiCr loads, calibrated RuOx thermometer, and a heater connected to a room-temperature proportional-integral-derivative (PID) loop for precise temperature control. Two loads were connected to the RF switch at reference planes A and B via NbTi superconducting cables, which ensured low-loss signal transmission while maintaining weak thermal coupling to the mixing chamber. Cryogenic directional couplers, Quantum Microwave QMC-CRYOCOUPLER-16, were used to inject probe signals and pump tones into the mTWPA input, as well as to extract probe signals from the output for characterization.

Input and output signals from the dilution refrigerator were routed to a vector network analyzer (VNA) and a spectrum analyzer via an RF switch bank. The VNA forward probe signal and the primary pump tone $\omega_{p0}$ were combined at room temperature using a directional coupler and routed through RF switches to both input ports of the dilution refrigerator, facilitating the characterization of $S_{21}$, $S_{12}$, and $S_{11}$. Additional RF sources at $\omega_{p1}$ and $\omega_{p2}$ were used to inject a forward pump during $S_{12}$ measurements and a variable reverse signal $P_{\rm{rev}}$ during $S_{21}$ measurements, respectively. The VNA was calibrated at reference planes A and B at cryogenic temperatures, and it was utilized to measure gain, return loss, isolation, and the dynamic range of the mTWPA as functions of pump frequency, pump power, signal power, and $\Phi/\Phi_0$. The spectrum analyzer was used to measure the signal and noise power levels of both HEMT amplifier chains, thereby characterizing mTWPA gain and system noise power.

\section{\label{AppendixD}mTWPA Implementation}

\begin{figure}[t]
\includegraphics[scale=0.8]{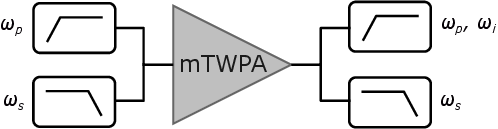}
\caption{\label{figA5} Practical implementation of the mTWPA with diplexers to separate the signal and the idler and pump onto physically separate ports to prevent back-action on the quantum device under measurement and potential saturation of the amplifier readout chain.}
\end{figure}

The integration of the mTWPA into a practical measurement setup can be realized with diplexers \cite{twpaperiphckts1}. Diplexers utilize high-pass and low-pass filters to separate the signal and pump at both the input and output of the mTWPA, as illustrated in Figure \ref{figA5}. Separation of the signal and the pump is crucial to preventing unwanted reflections of the pump tone from the input port of the mTWPA, which could otherwise interfere with the quantum device under measurement. Efficient separation of signal and pump mitigates the risk of the strong pump propagating through the mTWPA and saturating the readout amplifier chain at the output port. A key aspect of the diplexer design is ensuring that the high-pass and low-pass filters exhibit a sharp roll-off and are designed to be singly terminated. This guarantees that the common port, interfacing with both the input and output ports of the mTWPA, presents a well-matched impedance across the entire operating frequency range of the diplexer. 

\section{\label{AppendixE}WRSpice Simulation}

\begin{figure*}[t]
\includegraphics[scale=1.2]{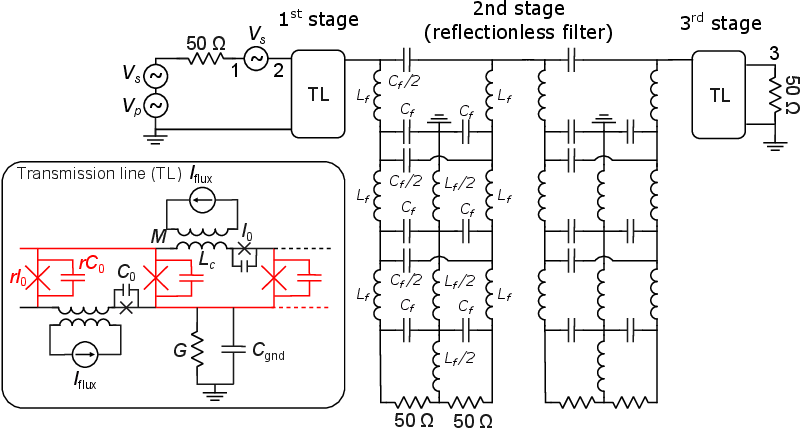}
\caption{\label{figA6} Circuit model utilized for time-domain WRspice simulations of the mTWPA. The first and third stages are transmission lines (TL) modeled as a ladder network of coupled asymmetric SQUIDs shown in inset. The circuit parameters are described in the text. The reflectionless filter is modeled with a lumped-element filter topology. Two stages of the reflectionless filter are used to provide sufficient filter roll-off and stop-band isolation. For extraction of S-parameters, a stimulus circuit composed of three voltage sources was used at the input to the first stage of the mTWPA. Extraction of S-parameters is accomplished by simulating the voltages at nodes 1-3.}
\end{figure*}

WRspice circuit simulations were used to model the gain characteristics of the mTWPA \cite{wrspce_pub,wrspice,twpasim1,twpasim2,twpasim3}. The circuit schematic utilized for these simulations is shown in Figure \ref{figA6}. The nonlinear transmission line (TL) segments that make up the first and third stages of the mTWPA comprise a ladder network of coupled asymmetric SQUID unit cells. Each unit cell consists of small and large Josephson junctions with critical currents $I_0$ and $rI_0$ and shunting capacitance $C_0$ and $rC_0$ respectively. The unit cell includes a ground capacitance $C_{\rm gnd}$ and a conductance $G$ to account for losses. The SQUIDs are flux biased using a DC current $I_{\rm flux}$ through a mutual inductance $M$ to the SQUID loops. The inductance contribution of the SQUID loop coupling inductor is negligible in comparison to the junction inductance within the SQUID loops. Each transmission line segment in the simulation consists of 350 unit cells, consistent with the mTWPA characterized in experiment.

The balanced high-pass reflectionless filter in the experimental device was realized using two quadrature hybrids and multi-pole high-pass filters. Due to the complexity of accurately modeling this filter as a lumped-element equivalent circuit in WRspice, an alternative approach was adopted. The reflectionless filter was approximated using the lumped-element design by M. Morgan \cite{morgan}. A two-stage lumped-element topology of the reflectionless filter was used to approximate the cutoff frequency, stop-band isolation, and roll-off characteristics observed in experiment as shown in \ref{figA6}. The inductance $L_f$ and capacitance $C_f$ values of the reflectionless filter used in the simulation were 1.4 nH and 0.7 pF, respectively. 

Time-domain WRspice simulations were used to simulate the forward signal gain, as presented in Figure \ref{fig4}(a) (red dashed line) which is in good agreement with experiment. The pump and signal were introduced as sinusoidal voltage sources with amplitudes $V_p$ and $V_s$, respectively. In this configuration, $S_{21}$ was determined from the simulated voltage amplitudes $V_n$ at node $n=3$ shown in Figure \ref{figA6} using the relation $S_{21} = V_3/V_s$ at the signal frequency $\omega_s$. To correlate the WRspice simulations with experimental data, simulations of $\angle S_{21}$ as a function of pump voltage $V_p$ were performed and compared to the experimental transmission phase measurements at the optimal operating point in $P_p$ of the mTWPA. 

WRspice simulations were used to compute $S_{11} = V_1/V_s$ and the total voltage sum of the forward and reverse traveling waves at the input $V_2$. Simulated return loss results for different values of first-stage $l$ and $Z_f$ are shown in Figures \ref{figA1} and \ref{figA2}.

WRspice time-domain simulations were used to compute the noise at the input of the mTWPA. Gaussian noise was introduced with a $V_{\rm rms}(\omega_s) = \sqrt{50 \times BW \times \hbar \omega_s/2 \coth \left(\hbar \omega_s/2k_BT \right)}$
where $BW = 50$ GHz and a temperature $T = 10$ mK. The generated noise data was imported as a piecewise linear voltage source for $V_s$ in WRspice. The backward-propagating noise and total noise at the input of the mTWPA were simulated at nodes $V_1$ and $V_2$ respectively. Noise simulations were conducted for various first stage $l$ and $Z_f$, as shown in Figures \ref{figA1} and \ref{figA2}. The noise simulations did not incorporate additional noise contributions from losses inherent to the mTWPA which is beyond the scope of this work.

\section{\label{AppendixF}Noise Measurement Calibration}

Noise measurements of the mTWPA were conducted using the signal-to-noise ratio (SNR) improvement method. The initial step in characterizing the mTWPA noise involved measuring the effective noise of the HEMT amplifier readout chain. In our experimental setup, two readout chains needed to be characterized, determining both gain and effective noise at reference planes A and B, as shown in Figure \ref{figA4}.

A broadband thermal noise source, consisting of two 50 $\Omega$ loads connected to reference planes A and B was varied in temperature between 150 mK and 4 K. This setup was utilized to perform a Y-factor measurement \cite{pozar}. The noise power was measured using a spectrum analyzer, which was switched between the outputs of the two amplifier readout chains at room temperature, as shown in Figure \ref{figA4}. The measured noise power was analyzed and fitted to the following noise model:

\begin{eqnarray}\label{eq12}
P_{\rm noise}(\omega, T) = \nonumber\\
\left[\frac{\hbar\omega}{2}\left(\text{coth}\frac{\hbar\omega}{2k_BT}\right)+k_B T_{\rm HEMT}(\omega)\right]G(\omega)B,
\end{eqnarray}

\noindent where $T_{\rm HEMT}(\omega)$ is the effective noise temperature of either HEMT A or B, $G(\omega)$ is the system gain, and $B$ is the measurement bandwidth. The first term in Eq. \ref{eq12} represents the noise emitted from the noise source at frequency $\omega$ when heated to temperature $T$. The measured $T_{\rm HEMT}(\omega)$ represents the effective noise temperature of the readout chain, including contributions from the HEMT amplifier and additional losses before the HEMT, originating from cables and isolators used in the experimental setup. The effective noise temperatures $T_{\rm HEMTA}(\omega)$ and $T_{\rm HEMTB}(\omega)$ was determined for reference planes A and B respectively.

\clearpage

\bibliography{twpaisolator}

\end{document}